%% file: main.tex
\documentclass[pmlr]{jmlr}
\usepackage{wrapfig}
\usepackage{adjustbox}
\usepackage{adjustbox}
\usepackage{booktabs}
\usepackage{multirow}
\usepackage{listings}
\usepackage{color}
\usepackage[most]{tcolorbox}

\usepackage{soul}

\colorlet{soulblue}{blue!20}

\usepackage{xcolor} 
\lstdefinelanguage{json}{
    basicstyle=\ttfamily\footnotesize,
    numbers=left,
    numberstyle=\tiny,
    stepnumber=1,
    showstringspaces=false,
    breaklines=true,
    frame=single,
    backgroundcolor=\color{gray!10},
    keywordstyle=\color{jsonkey},
    stringstyle=\color{jsonstring},
    morestring=[b]",
    morecomment=[l]{//},
    morecomment=[s]{/*}{*/},
    morekeywords={true,false,null}
}
\usepackage[normalem]{ulem}

\hypersetup{
	colorlinks=true,
	linkcolor=black,
	urlcolor=black,
	linktoc=all
}
\RequirePackage{graphicx}
 \usepackage{booktabs}
\usepackage{longtable}
\makeatletter
\def\set@curr@file#1{\def\@curr@file{#1}} 
\makeatother
\usepackage[load-configurations=version-1]{siunitx} 

\theorembodyfont{\upshape}
\theoremheaderfont{\scshape}
\theorempostheader{:}
\theoremsep{\newline}

\jmlrvolume{[VOLUME \# TBD]}
\jmlryear{2026}
\jmlrworkshop{Machine Learning for Healthcare}

\author{%
  \Name{Gaia A. Bertolino}
  \Email{gab62@cam.ac.uk}\\
  \addr University of Cambridge, United Kingdom
  \AND
  \Name{Yuwei Zhang}
  \Email{yz798@cam.ac.uk}\\
  \addr University of Cambridge, United Kingdom
  \AND
  \Name{Tong Xia}
  \Email{tongxia@mail.tsinghua.edu.cn}\\
  \addr Tsinghua University, China
  \AND
  \Name{Domenico Talia}
  \Email{talia@dimes.unical.it}\\
  \addr University of Calabria, Italy
  \AND
  \Name{Cecilia Mascolo}
  \Email{cm542@cam.ac.uk}\\
  \addr University of Cambridge, United Kingdom
}
\title[RAMoEA-QA]{RAMoEA-QA: Hierarchical Specialization for Robust Respiratory Audio Question Answering}
\begin{document}

\maketitle

\begin{abstract}
Conversational generative AI is increasingly explored in healthcare, where models must integrate heterogeneous patient signals and support diverse interaction styles while producing clinically meaningful outputs. In respiratory care, non-invasive audio recordings captured with sensing devices offer a scalable route to screening and longitudinal monitoring, but heterogeneity is particularly acute: recordings vary across devices, environments, and acquisition protocols, and queries may vary in intent, answer format, and prediction objective
. Existing biomedical audio-language question answering systems for respiratory assessment are starting to emerge, 
but they are typically built as single-path models, processing all inputs through the same acoustic and language pathway despite variation in recording conditions and query types. 
They are also usually evaluated in relatively limited settings, leaving open their robustness under realistic distribution shifts, including changes in acquisition domains, modality, and clinical task.


To address this gap, we introduce \textbf{RAMoEA-QA}, the first respiratory audio question answering model designed to support input-dependent specialization across heterogeneous recordings and query types within a unified hierarchical two-stage framework. 
We study this design in a unified respiratory audio question answering setting spanning clinical and self-recorded, multi-device acquisition settings, question formats, and both discrete and continuous targets. Across in-domain and controlled-shift evaluations, RAMoEA-QA improves over matched monolithic baselines and routing controls, reaching $0.72$ in in-domain test accuracy (vs.\ $0.61$ and $0.67$ for single-path baselines) on discriminative tasks, while also achieving the best regression performance and stronger average transfer under dataset, modality, and task shifts, including gains of up to 23 percentage points in accuracy on the COPD modality-shift setting.
\end{abstract}

\input{sections/1_introduction}
\input{sections/2_related_work}
\input{sections/3_method}
\input{sections/4_experimental_setup}
\input{sections/5_results}
\input{sections/6_conclusions}
\bibliography{references}
\appendix
\input{appendix/datasets_overview}
\input{appendix/metrics_overview}

\input{appendix/method}

\input{appendix/baselines}
\input{appendix/example}
\input{appendix/results}

\end{document}

%% file: sections/1_introduction.tex
\section{Introduction}
\label{sec:intro}

Respiratory diseases remain a leading cause of global morbidity and mortality, motivating early, accessible, and scalable screening tools, especially for telemedicine and low-resource settings~\citep{article}. In clinical practice, respiratory sound assessment provides diagnostic cues (e.g., wheezes, crackles, timing patterns) for conditions such as asthma, pneumonia, and chronic obstructive pulmonary disease (COPD). Recent machine learning work has achieved strong results for respiratory sound classification from recordings and has enabled reusable pretrained audio backbones for downstream respiratory tasks~\citep{Reichert2008, article_3, zhang2024openrespiratoryacousticfoundation}.

Despite the progress, existing systems typically produce a single, task-specific output (e.g., disease label or abnormality detection), which limits their usefulness in real clinical interactions, where the same recording may need to support multiple conversational turns and different follow-up questions about diagnosis, severity, symptoms, or physiological measurements (e.g., \emph{Is wheezing present?}, \emph{Which diagnosis is most likely?}, \emph{How severe is it?}, \emph{What is the respiratory rate?}). 
Enabling flexible question answering over respiratory audio would allow \emph{interactive} respiratory assessment beyond a single, static prediction. Although general-purpose multimodal assistants are increasingly explored for conversational healthcare support~\citep{munikoti2024generalist,openai_chatgpt_health,anthropic_healthcare_ls,google_health_ai}, and clinical question answering (QA) has been widely studied for modalities such as electronic health records, medical imaging, and biosignals~\citep{pampari2018emrqalargecorpusquestion, nguyen_medredqa_2023, oh_ecg-qa_2023}, conversational AI for respiratory audio remains under-explored.

Existing large-scale audio-language foundation models~\citep{deshmukh_pengi_2024,das_speechverse_2024, chu_qwen-audio_2023}  have proven to perform well for 
generic audio understanding, yet they are unreliable for respiratory assessment~\citep{Wang_Chen_Zeghidour_Saeed_2025}. This limitation stems from the lack of training on respiratory acoustics and clinical knowledge, where cues are subtle, recordings are uncontrolled, and sound-label relationships are noisy and context-dependent.
Early biomedical audio-language QA models partially address these limitations but are developed and validated in relatively constrained settings (e.g., limited question formats and prediction objectives), and rarely study how to specialize across diverse respiratory corpora and query intents, which is crucial because real-world respiratory audio question-answering must generalize across question formats, \emph{modality} (cough/breath/speech), \emph{dataset/device}, and \emph{clinical task} shifts.
It therefore remains unclear whether
a single monolithic framework 
can reliably support the diversity of clinically structured questions encountered in practice~\citep{Wang_Chen_Zeghidour_Saeed_2025}. 

To address these limitations, we introduce \textbf{RAMoEA-QA}, to our knowledge the first respiratory audio question answering model to apply \textbf{hierarchical two-stage conditional specialization} across both audio encoding and language adaptation
. The design is motivated by the two main sources of heterogeneity in respiratory audio question answering (RA QA): variability in the \emph{acoustic input} (e.g., recording conditions, devices, and respiratory modalities) and variability in the \emph{query space} (e.g., question intents, answer formats, and prediction objectives). RAMoEA-QA addresses these through hierarchical specialization across both the audio and language components, using an \textbf{Audio Mixture-of-Experts} to select the most suitable encoder for each recording and a \textbf{Language Mixture-of-Adapters} to adapt generation to the question and expected answer style. This yields a unified multimodal model that can adapt to realistic clinical variability without requiring all branches to be active for every example.


We study this design on the \textbf{RA-QA collection}~\citep{Bertolino_Zang_Xia_Talia_Mascolo_2026}, a multi-dataset respiratory audio QA benchmark spanning diverse acquisition regimes, including both clinical and self-recorded settings, three question formats (\emph{open-ended}, \emph{single-verify}, and \emph{multiple-choice}), and two task families (\emph{discriminative} and \emph{regression}). Beyond standard in-distribution evaluation, we conduct controlled robustness tests under \textbf{modality}, \textbf{dataset}, and \textbf{task} shifts, analyze routing behaviour to characterize specialization patterns and evaluate the effects of routing collapse through fixed-path controls, and perform ablation studies on the proposed framework. In this way, RAMoEA-QA serves not only as the first hierarchically specialized model for respiratory audio question answering, but also as a testbed for understanding how conditional specialization performs in a heterogeneous clinical audio QA setting.
Our main contributions are as follows:
\begin{itemize}
    \item We introduce \textbf{RAMoEA-QA}, to our knowledge the first respiratory audio QA model with \textbf{hierarchical two-stage routing} over both audio experts and language adapters, enabling example-wise specialization across heterogeneous recordings, question formats, and prediction objectives while preserving efficient single-path inference at test time with limited additional parameters.
    
    \item We study respiratory audio QA in a unified evaluation setting on the RA-QA collection, covering \textbf{multiple datasets}, \textbf{multiple question formats}, \textbf{discriminative and regression targets}, and controlled \textbf{modality}, \textbf{dataset}, and \textbf{task} shifts.
    
    \item We show that conditional specialization is a competitive and robust design choice for RA QA in our setting: compared with matched monolithic baselines and routing controls, RAMoEA-QA improves overall in-domain performance and shows stronger average transfer in the evaluated shift settings, while routing analysis reveals structured specialization patterns rather than uniform use of a single path.
\end{itemize}

\subsection*{Generalizable Insights about Machine Learning in the Context of Healthcare}
While our results are specific to respiratory audio question answering, they suggest several lessons for machine learning on heterogeneous clinical signals. First, when both inputs and queries are heterogeneous, \emph{conditional specialization} can be more effective than forcing all examples through a single shared pathway. Second, useful specialization does not necessarily require fully separate end-to-end models: routing over \emph{frozen domain backbones} combined with \emph{parameter-efficient language adaptation} can improve robustness while preserving efficient single-path inference. Third, evaluation of clinical QA systems should distinguish \emph{task correctness} from \emph{text-level response quality}, since fluent answers may still be clinically uninformative or incorrect.

%% file: sections/2_related_work.tex
\section{Related Work} 
\label{sec:related_work}

\paragraph{Audio ML for respiratory health.}
Machine learning for respiratory acoustics has primarily focused on supervised classification and detection using cough and breath recordings, targeting tasks such as disease screening, symptom identification, and severity estimation~\citep{9630091, article1, article2, Reichert2008, article_3}. More recently, respiratory \emph{audio foundation models} have been introduced to leverage large-scale pretraining for improved transfer across datasets and recording conditions~\citep{zhang2024openrespiratoryacousticfoundation}. While these approaches provide strong acoustic representations and competitive performance on standard prediction tasks, they are typically designed for fixed outputs (labels or scores) and do not directly support question-conditioned clinical interactions where the same recording may be queried in multiple ways (e.g., diagnosis, symptom verification, or severity/regression targets).
\paragraph{QA systems in healthcare.}
Clinical question answering has been extensively studied in text-centric settings, including QA over clinical notes and biomedical literature~\citep{pampari2018emrqalargecorpusquestion, jin_pubmedqa_2019, nguyen_medredqa_2023, Tsatsaronis_Balikas_Malakasiotis_Partalas_Zschunke_Alvers_Weissenborn_Krithara_Petridis_Polychronopoulos_etal._2015}, as well as structured EHR query answering~\citep{wang2020texttosqlgenerationquestionanswering, lee2023ehrsqlpracticaltexttosqlbenchmark}. Beyond text, multimodal medical QA benchmarks have emerged for modalities such as medical imaging and biosignals~\citep{abacha_vqa-med_nodate, hu_omnimedvqa_2024, oh_ecg-qa_2023}, alongside approaches that integrate large language models with modality-specific encoders for instruction following and multimodal reasoning~\citep{singhal2022largelanguagemodelsencode, wu2024nextgptanytoanymultimodalllm}. More broadly, recent work emphasizes a shift toward \emph{generalist} assistants that aim to handle diverse modalities, tasks, and interaction styles within a single interface, including emerging healthcare-facing conversational systems~\citep{munikoti2024generalist,openai_chatgpt_health,anthropic_healthcare_ls,google_health_ai}.

In the audio domain, general-purpose audio-language models have demonstrated broad capabilities across tasks~\citep{deshmukh_pengi_2024, das_speechverse_2024, chu_qwen-audio_2023, gong2024listenthinkunderstand, ghosh2024gamalargeaudiolanguagemodel, huang2023audiogptunderstandinggeneratingspeech}. More recently, respiratory-focused resources such as CareAQA~\citep{Wang_Chen_Zeghidour_Saeed_2025} have targeted biomedical audio question answering, but they typically emphasize a narrower set of respiratory objectives, such as only one question type, rather than robust, multi-format clinical querying. This leaves a gap between generalist conversational systems and clinically grounded respiratory audio QA: systems for respiratory audio QA in practice must handle heterogeneous question styles (open-ended explanations, verification, multiple-choice) and task families (discriminative and regression-style severity targets) under dataset and recording-condition shift. Prior respiratory audio QA systems remain limited in scale and scope, and they rarely study conditional specialization across heterogeneous respiratory data and query types.
\paragraph{MoE/MoA architectures for specialization, efficiency, and robustness.}
Mixture-of-Experts (MoE) enables conditional computation by routing each input to a subset of experts, improving scalability and capacity while controlling computation~\citep{jacob, lepikhin2020gshardscalinggiantmodels, fedus2022switchtransformersscalingtrillion}. In parallel, parameter-efficient fine-tuning methods such as adapters and LoRA have made it practical to adapt large pretrained models without updating all parameters~\citep{Pfeiffer_Kamath_Ruckle_Cho_Gurevych_2021, hu2021loralowrankadaptationlarge}. Building on these ideas, mixture-of-adapters (often referred to as MoA) train multiple lightweight adapters that are dynamically selected at inference time, enabling modular specialization with limited trainable parameters~\citep{Lee_Jang_Kim_Jung_Kim_2024, Buehler_Buehler_2024, Wu_Huang_Wei_2024, Wang_Li_2024}. Beyond efficiency, conditional specialization can improve \emph{robustness under distribution shifts} by activating experts aligned to the current input domain (e.g., device, modality, cohort) or query formats, rather than forcing a single parameterization to fit all conditions~\citep{Guo_Shah_Barzilay_2018}.
A known challenge in mixture-based systems is \emph{routing imbalance} and collapse, where a small subset of experts receives most of the traffic and parameters are effectively under-utilized; prior work has studied load-balancing, entropy regularization, and related stabilization strategies in sparse mixtures~\citep{fedus2022switchtransformersscalingtrillion, lepikhin2020gshardscalinggiantmodels}. Despite their application in Natural Language Processing~\citep{wu2024mixtureloraexperts}, have seen limited study as a joint specialization mechanism in multimodal clinical QA, and have not been studied for respiratory audio QA.

%% file: sections/3_method.tex
\section{Method}
\label{sec:method}

\input{figures/figures_tex/architecture}

This section describes \emph{RAMoEA-QA}, our unified framework for respiratory audio question answering that jointly adapts acoustic processing and language generation to heterogeneous recordings and queries.

\subsection{Overview} \label{sec:method_overview}
RAMoEA-QA is a unified multimodal architecture for respiratory audio question answering designed to adapt processing jointly to two sources of heterogeneity: variability in the respiratory recording and variability in the clinical query. To achieve this, the model uses two complementary specialization mechanisms that operate at different stages of the pipeline. On the audio side, an \textbf{Audio Mixture-of-Experts (Audio-MoE)} selects the most suitable pre-trained respiratory audio encoder for the input recording. On the language side, a \textbf{LoRA Mixture-of-Adapters (LoRA-MoA)} selects the most suitable LoRA adapter on a shared frozen LLM to match the question and expected answer style. Together, these components yield a hierarchical two-stage specialization mechanism within a single unified model. In the default configuration presented here, RAMoEA-QA uses two audio experts and two LoRA adapters.
Given an audio recording and a natural language question, RAMoEA-QA generates a short answer whose \emph{format and prediction objective} are determined by the query, allowing a single system to support multiple question formats and both discrete and continuous targets for the same audio recording. For each example, the model activates exactly \emph{one} audio expert and exactly \emph{one} LoRA adapter through hard top-1 routing, enabling input-dependent specialization while retaining single-path inference at test time.

\figureref{fig:architecture} summarizes the full pipeline and highlights the two routing stages: \textbf{(A) Audio-MoE} and \textbf{(B) LoRA-MoA}. In stage (A), a lightweight routing representation is computed from the input and used to select one audio expert. The selected expert encodes the recording, and its output is aligned to the LLM hidden size and injected as a \emph{selected audio prefix} (aligned audio embeddings concatenated into the LLM input). In stage (B), conditioned on the question and the selected aligned audio prefix, the model selects one LoRA adapter from the LoRA-MoA and generates the answer using the shared frozen language backbone.

\subsection{Routing configuration}
In our setting, \emph{routing} denotes the expert selection mechanism: which audio encoder and which language adapter are activated for a given input pair. Each router computes gating logits from a lightweight routing representation derived from the available input at its stage and uses them to select a single expert, as described in Section~\ref{sec:method_overview}.

Routing only determines which expert pathway is selected; it does not alter the specific inputs provided to the experts themselves as audio encoders always receive the spectrogram, and the LLM always receives the prompt together with the audio prefix.

\subsection{Audio Mixture-of-Experts (Audio-MoE)}
\label{sec:method_audio_moe}
\paragraph{Lightweight routing proxy.}
As illustrated in the \textcolor{orange}{orange} block in Fig.~\ref{fig:architecture}, the router assigns the input to one of the available experts.
To avoid running all audio encoders, the Audio-MoE first computes a low-cost \emph{routing proxy} from the spectrogram and the question, and uses it to select a single audio expert.

\paragraph{Expert activation and alignment.}
Let $\{\mathcal{E}^{a}_{1},\ldots,\mathcal{E}^{a}_{N_a}\}$ be heterogeneous pre-trained audio encoders used as \emph{frozen} foundation backbones. After the router selects an expert, only the chosen encoder processes the spectrogram, producing an embedding sequence that is mapped to the LLM hidden size via an expert-specific aligner and injected into the LLM input as a \emph{selected audio prefix} (aligned audio embeddings concatenated into the LLM context). 

\subsection{Language Mixture-of-Adapters (LoRA-MoA)}
\label{sec:method_moa}
\paragraph{Shared LLM with adapter experts.}
As illustrated in the \textcolor{green}{green} block in Fig.~\ref{fig:architecture}, we use a shared frozen autoregressive language backbone and attach $N_\ell$ LoRA adapters as lightweight language experts. The MoA router selects exactly one adapter per example, enabling specialization across question formats and task families while keeping the LLM backbone fixed.

\paragraph{Router input without re-embedding.}
After audio expert selection and alignment, the MoA router computes its routing vector using the same \texttt{fused} router-input policy instantiated at the language stage, performing a lightweight fusion between the question representation and the aligned audio embeddings.

\paragraph{Complementary two-stage fusion.}
Importantly, the two routers do \emph{not} condition on the same signal: (A) fuses the question with a \emph{cheap pre-encoder proxy} \emph{before} selecting an audio expert, whereas (B) fuses the prompt with \emph{expert-produced aligned audio embeddings} \emph{after} expert selection. This staged conditioning makes routing decisions complementary (coarse acoustic/domain choice vs.\ generation/style/intent refinement), rather than redundant.

\subsection{Training objective and routing regularization.}
We train with teacher forcing and compute the next-token negative log-likelihood only over answer tokens, masking out prompt and audio-prefix positions in the loss. Let $\mathcal{L}_{\text{main}}$ denote this answer-masked causal language modeling loss. For routing, let $\mathbf{p}_a\in\mathbb{R}^{N_a}$ and $\mathbf{p}_\ell\in\mathbb{R}^{N_\ell}$ be the softmax routing probabilities over the $N_a$ audio experts and $N_\ell$ LoRA adapters, and let $\mathbf{u}_a,\mathbf{u}_\ell$ be the corresponding hard one-hot assignments (from straight-through Gumbel-Softmax during training). We add load-balancing regularization $\mathcal{L}_{\text{LB}}(\cdot)$ to encourage non-degenerate expert utilization, and optionally an entropy term $\mathcal{L}_{\text{ENT}}(\cdot)$ to control distribution sharpness. The full objective is:
\begin{equation}
\begin{aligned}
\mathcal{L} &= \mathcal{L}_{\text{main}}
+\lambda_a\,\mathcal{L}_{\text{LB}}(\mathbf{p}_a,\mathbf{u}_a)
+\lambda_\ell\,\mathcal{L}_{\text{LB}}(\mathbf{p}_\ell,\mathbf{u}_\ell) \\
&\quad -\beta_a\,\mathcal{L}_{\text{ENT}}(\mathbf{p}_a)
-\beta_\ell\,\mathcal{L}_{\text{ENT}}(\mathbf{p}_\ell),
\end{aligned}
\end{equation}
where $\lambda_a,\lambda_\ell\ge 0$ weight the load-balancing terms for the audio and language routers, and $\beta_a,\beta_\ell\ge 0$ weight the entropy regularizers.
More implementation details are reported in \appendixref{apd:method_appendix_math}.

%% file: figures/figures_tex/architecture.tex
\begin{figure*}[htbp]
\floatconts
  {fig:architecture}
  {\caption{RAMoEA-QA architecture. (A) The MoE selects an audio expert (encoder). The resulting aligned audio embeddings are injected as a selected audio prefix. (B) The MoA selects a LoRA adapter for the language model during generation.}}
  {\includegraphics[width=\linewidth]{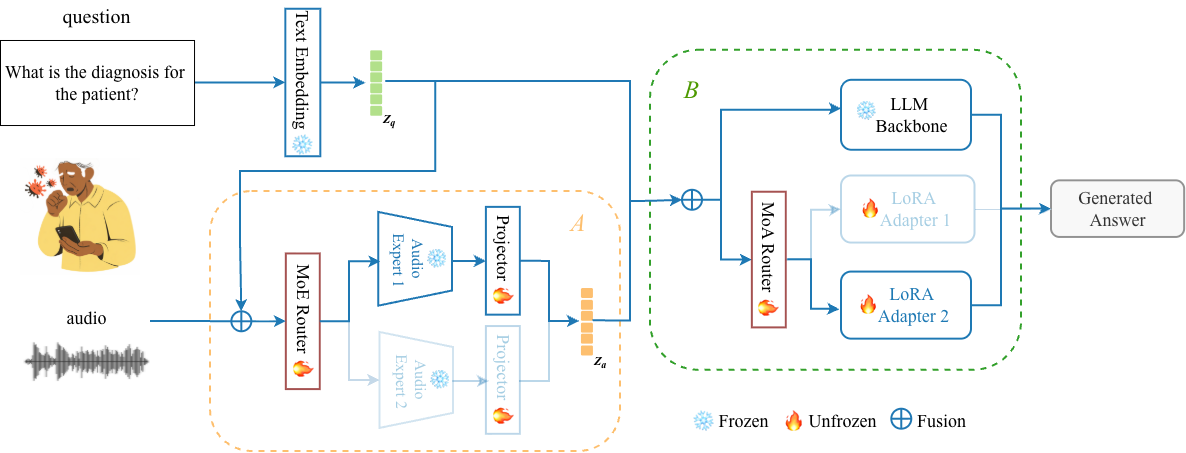}}
\end{figure*}

%% file: sections/4_experimental_setup.tex
\section{Experimental Setup} 
\label{sec:experimental_setup}

\input{tables/types_of_questions}
\subsection{Datasets and tasks.} 
We conduct all experiments on the \textbf{RA-QA collection}, which harmonizes multiple publicly available respiratory audio datasets into a unified question-answering setting. Our training mixture comprises seven datasets from the collection, spanning different acquisition settings, audio modalities, and labeling schemes (Table~\ref{tab:data_stats}). This setup allows us to study respiratory audio QA under heterogeneity in both input conditions and query structure, rather than within a single dataset or task formulation. To explicitly evaluate robustness beyond standard in-distribution testing, we further define controlled held-out settings along three axes (Table~\ref{tab:generalization}): (\emph{i}) \textbf{modality shift} ($\Delta M$), where we evaluate on audio modalities not used during training for a dataset (e.g., Coswara asthma trained on cough and tested on breathing/counting/vowel); (\emph{ii}) \textbf{dataset shift} ($\Delta D$), where an entire dataset is held out from training and used only at evaluation (e.g., UK COVID-19); and (\emph{iii}) \textbf{combined task and modality shift} ($\Delta TM$), where we evaluate on a held-out clinical task together with modalities not seen during training.

\paragraph{Task families.}
We organize questions into two task families: \textbf{Discriminative} (predicting a categorical respiratory condition), including presence of disease and \textit{Severity} (predicting an ordinal/discrete severity level), and \textbf{Regression} (predicting continuous physiological values such as spirometry measures, FEV1, FVC, FEV1/FVC, and respiratory rate).

\paragraph{Question formats}
\label{sec:q_formats}
We formulate respiratory assessment as \textbf{question-conditioned prediction} from audio. Each sample consists of a recording $x$ and a natural-language question $q$, and the model produces an answer $\hat{y}$ appropriate to the query. We evaluate three question formats, following~\citep{Bertolino_Zang_Xia_Talia_Mascolo_2026}:
\begin{itemize}
    \item \textbf{Open-ended (OE).} The model generates a free-form textual answer. This format is used for descriptive responses (e.g., likely condition, short clinically grounded summary) and for natural-language reporting of numeric targets.
    \item \textbf{Single-verify (SV).} The model answers a yes/no verification question (e.g., 'Does the patient suffer from asthma?'). We treat this as binary decision-making with standardized outputs (Yes/No) for evaluation.
    \item \textbf{Multiple-choice (MC).} Unlike open-ended QA, multiple-choice questions explicitly enumerate the candidate labels or options in the prompt, constraining generation to a closed set and making the task a \emph{selection} problem rather than free-form answering. 
\end{itemize}
For discriminative tasks, when a dataset provides multiple \emph{question formats} for the same recording (e.g., open-ended, single-verify, and/or multiple-choice), we treat each \emph{(audio, question)} pair as a separate example and include all available formats. 
For regression tasks, we only present questions in the open-ended format, since it is the most suitable. This formulation lets us study not only whether a model predicts the correct target, but also how its behaviour changes across answer formats and task types.

\subsection{Baselines.}
\label{sec:baselines}

We compare RAMoEA-QA against two classes of reference systems.

\paragraph{Monolithic QA baselines.}
Our primary baselines are CareAQA-style \emph{single-path} models~\citep{Wang_Chen_Zeghidour_Saeed_2025}, implemented as monolithic counterparts to our routed architecture. These models use one audio pathway and one LoRA adapter on a frozen LLM, and therefore provide the closest comparison for isolating the effect of hierarchical conditional specialization. We instantiate this baseline with two respiratory audio backbones, \textbf{OPERA-CT} and \textbf{OPERA-GT} from~\citep{zhang2024openrespiratoryacousticfoundation}.

\paragraph{Generic audio-language baselines.}
To contextualize respiratory audio QA within broader audio-language modelling, we additionally evaluate three generic audio-language models. \textbf{Pengi}~\citep{Deshmukh_Elizalde_Singh_Wang_2024} is included as a zero-shot reference that is relatively close in overall scale and framework, while \textbf{GAMA}~\citep{ghosh2024gamalargeaudiolanguagemodel} and \textbf{LTU}~\citep{gong2024listenthinkunderstand} are included as strong recent generic audio-language baselines reported in prior audio-QA work~\citep{Wang_Chen_Zeghidour_Saeed_2025}. 
These baselines are not specialized to respiratory assessment and are not finetuned on the task, thus they are used to assess out-of-domain transfer from generic audio-language models to respiratory QA.

\subsection{Evaluation Metrics}
\label{sec:metrics}
We follow the RA-QA protocol~\citep{Bertolino_Zang_Xia_Talia_Mascolo_2026} and report both \emph{text-level} and \emph{task-level} metrics to separately capture (i) how well the model matches the reference form and semantics, and (ii) whether it predicts the correct clinical label or value. 

\paragraph{Text-level metrics.} 
To evaluate answer quality at the word and semantic level, we compute 
\textbf{TokenF1} (token overlap), \textbf{BERTScore} (semantic similarity), and \textbf{METEOR} (lexical alignment) between generated and reference answers after standard normalization. These metrics reflect complementary aspects: TokenF1 emphasizes surface fidelity, METEOR captures word-level matching under minor morphological variation, and BERTScore provides a semantic view that is less sensitive to paraphrasing. We report overall averages and stratify by \texttt{question-type}.

\paragraph{Label-correctness metrics.}
For \textbf{discriminative tasks} (the diagnosis and severity task family), we evaluate \emph{label correctness} by extracting the target label from the model output and comparing it to the ground truth. For open-ended and multiple-choice questions, we parse the generated text to recover the predicted \emph{target label} (e.g., by normalizing case/punctuation and matching against the dataset label set), whereas for single-verify questions we parse the output into $\{\texttt{Yes},\texttt{No}\}$. We then compute \textbf{Macro-F1} and \textbf{Accuracy} over the extracted labels. 
For \textbf{regression tasks}, we extract the first valid numeric value from the generated answer (if present) and compare it to the ground-truth target. We report \textbf{MAE} and \textbf{RMSE} (lower is better) computed over successfully parsed examples. 

\subsection{Implementation details}
\label{sec:impl}
\paragraph{Audio preprocessing.} All models use the same audio preprocessing pipeline, following the OPERA~\citep{zhang2024openrespiratoryacousticfoundation} protocol for respiratory audio. Audio is resampled, trimmed to remove leading and trailing silence, and padded or truncated to a fixed duration; log-mel spectrograms are then computed using the OPERA parameters.

\paragraph{Routing training.}
Routing is trained with straight-through Gumbel-Softmax for expert exploration and evaluated at inference with argmax selection; we use a short balanced warmup to reduce routing collapse.

\paragraph{Optimization and decoding.}
We optimize with AdamW (learning rate $2\cdot 10^{-5}$, gradient clipping $1.0$) for up to 100 epochs (early stopping on the validation set). We train with a batch size of 4 and evaluate with a batch size of 1. 

\paragraph{Model setting (default).}
Our default configuration uses two OPERA audio experts (OPERA-CT and OPERA-GT) and a shared GPT-2 backbone with two LoRA adapters. GPT-2 was chosen as a lightweight shared backbone for a stable and computationally tractable setup, providing a controlled setting for studying the proposed framework.


\paragraph{Checkpoint selection.}
We select checkpoints based on the best in-distribution validation performance using the primary metric appropriate to the task family (Macro-F1 for discriminative; MAE for regression).

%% file: tables/types_of_questions.tex
\begin{table*}[hbtp]
\tiny
\floatconts
  {tab:data_stats}
  {\caption{\textbf{Overview of the datasets }used in our training, grouped by task family, with their target tasks, available audio modalities, and supported question formats.  }}
\centering
  {\begin{tabular}{llllc}
  \toprule
\bfseries Task Family & \bfseries Dataset & \bfseries Task  & \bfseries Audio condition & \bfseries Query-types   \\ \midrule
\multirow{5}{*}{Discriminative}  & Coswara& Asthma   & Deep and shallow breathing  & SV\\
   & COUGHVID  & Diagnosis (Healthy, Symptomatic, Covid) & Coughing  & All \\
   & ICBHI  & Diagnosis (URTI, COPD)  & Four clinical recording devices & All \\
   & KAUH   & Diagnosis (Healthy, LRTI, COPD)& Sound types and positions & All \\
   & Respiratory@TR& COPD Severity& Recording positions  & OE, MC \\ \midrule
\multirow{3}{*}{Regression}  & MM-Lung& Spirometry (FEV1)& Deep breathing   & OE   \\
   & MM-Lung& Spirometry (FVC) & Deep breathing   & OE   \\
   & Nosemic& Respiratory rate & Before and after running& OE   \\ \midrule
\end{tabular}}
\end{table*}

%% file: sections/5_results.tex
\section{Results}
\label{sec:results}

\input{tables/main_results}
We first compare RAMoEA-QA against general audio-language models 
and  monolithic baselines on in-distribution RA QA performance across question formats and task families. We then analyze routing behaviour and, finally, we study robustness under controlled modality, dataset, and task shifts, and report additional ablations and scaling experiments on the number of experts and adapters. Appendix~\ref{apd:qualitative_examples} reports a qualitative example of one discriminative and one regression task.
\subsection{Main results}
\label{sec:main_results}
\tableref{tab:main_results} summarizes overall performance across discriminative and regression task families, together with global text-level response metrics, for our model and all baselines. Specifically, we compare three generic audio-language models, \textbf{Pengi}, \textbf{LTU}, and \textbf{GAMA}, against monolithic single-path respiratory audio-QA baselines (\textbf{CaReAQA-style} with OPERA-CT and OPERA-GT) and our proposed \textbf{RAMoEA-QA}. For LTU, we found that the default prompting often led the model to abstain from answering; we therefore report results of an \textbf{LTU*} variant with an explicit instruction to always provide an answer.

\paragraph{Discriminative tasks.} 
Generic audio-language models remain limited for respiratory audio QA. As shown in \tableref{tab:main_results}, \textbf{Pengi} performs poorly ($0.22$ Accuracy, $0.21$ Macro-F1), while \textbf{LTU*} performs even worse on task-level correctness ($0.18$ Accuracy, $0.11$ Macro-F1). The stronger generic baseline \textbf{GAMA} improves over both ($0.45$ Accuracy, $0.28$ Macro-F1), but still remains well below respiratory-domain-aware models. This suggests that generic audio-language competence alone does not transfer reliably to clinically structured respiratory QA.

Within models finetuned to the RA-QA collection, our hierarchical two-stage routing improves discriminative performance over single-path baselines. Our model increases Macro-F1 from $0.53$ (OPERA-CT plus 1 LoRA adapter) and $0.59$ (OPERA-GT plus 1 LoRA adapter) to $0.67$, while also improving Accuracy (from $0.61$ and $0.67$ to $0.72$). 

\paragraph{Regression tasks.}
Generic audio-language models also struggle on regression tasks. \textbf{Pengi} does not yield parseable numeric outputs, resulting in no valid regression predictions. When not abstaining, \textbf{LTU*} produces parseable numbers with very poor accuracy (MAE $14.36$, RMSE $36.20$), indicating that forcing an answer is not sufficient for reliable numerical estimation. \textbf{GAMA} performs substantially better than the other generic models and yields consistent numeric outputs, but still remains worse than respiratory-domain-aware models, with an MAE of $2.83$ and RMSE of $4.21$. Among respiratory audio QA models, our two-stage routing mechanism further improves regression performance: while the best single-path baselines reach an MAE of $2.61$ and an RMSE between $3.93$ and $4.26$, RAMoEA-QA reduces error to $2.29$ MAE and $3.77$ RMSE.

\input{figures/figures_tex/tolerance}

Following common practice in biomedical measurement validation~\citep{OBrien_Waeber_Parati_Staessen_Myers_2001, Elgendi_Haugg_Fletcher_Allen_Shin_Alian_Menon_2024}, we complement MAE and RMSE with tolerance accuracy Acc@$\tau$, i.e., the cumulative fraction of predictions whose absolute error falls within a clinically meaningful threshold $\tau$ .
\figureref{fig:tolerance} shows that two-stage routing attains higher Acc@$\tau$ at \emph{stricter} tolerances on multiple spirometry and physiology targets, while converging to similar performance at looser tolerances. Together, these results indicate that conditional specialization improves both average regression quality and reliability under clinically relevant error budgets.

\paragraph{Text-level response correctness.}
Beyond task-level correctness, \tableref{tab:main_results} shows that RAMoEA-QA also produces the strongest responses at the semantic and lexical level. It achieves the highest global \textbf{BERTScore} ($0.90$) and \textbf{METEOR} ($88.38$), improving over both matched single-path baselines (e.g., $0.89/87.05$ for CareAQA-GT) and substantially outperforming generic audio-language models such as \textbf{GAMA} ($0.76/70.63$) and \textbf{Pengi} ($-0.01/2.61$). Token-level agreement also increases (from $0.83$ and $0.86$ of monolithic baselines to $0.88$), indicating more faithful short-answer generation. Interestingly, \textbf{LTU*} attains a high BERTScore ($0.90$) despite low task-level correctness. Inspection of its outputs shows that LTU often abstains from giving a grounded answer, thus remaining semantically fluent while still being clinically uninformative. These results indicate that the benefit of hierarchical specialization is not limited to label extraction or numerical prediction, but extends to the quality and alignment of the generated answers themselves, and that semantic-overlap metrics alone are not necessarily aligned with answer quality in structured clinical QA.

\paragraph{Performance across question types and datasets.} 
RAMoEA-QA is overall the strongest model at both the task and text level across question-types, with the best performance for \texttt{single-verify}, where it improves Macro-F1 to $0.82$ (vs.\ $0.62$ for the strongest monolithic baseline) and BERTScore to $0.94$ (vs.\ $0.91$), and \texttt{multiple-choice}
, with a small trade-off on \texttt{open-ended} questions. The per-dataset analysis confirms that RAMoEA-QA is the strongest model, with some trade-offs in more challenging settings.
Full per-question-type results are reported in  the appendix.

\subsection{Routing analysis}
\label{sec:routing_analysis}
To understand \emph{why} and \emph{how} routing helps, we analyze empirical routing distributions for our best two-stage model. 
\figureref{fig:general_routing} reports the fraction of samples routed to each audio expert (OPERA-CT vs.\ OPERA-GT) and each language expert (LoRA adapter 1 vs.\ 2), stratified by dataset, question format, and task.

\input{figures/figures_tex/general_routing}

\paragraph{Recording-condition-aware specialization.}
Routing is non-uniform: the model prefers different experts consistently with recording-condition and acquisition-aware specialization. For example, in Coswara dataset, matched deep- versus shallow-breath recordings from the same subject switch route in about 20\% of cases, while in ICBHI, recordings from the same patient and device family are assigned to different experts in 50\% of cases. These patterns suggest that routing adapts to within-dataset variation in acquisition and signal characteristics, rather than merely partitioning samples by corpus.


\paragraph{Coupled but non-redundant two-stage routing.}
A notable pattern is that adapter routing largely mirrors audio routing (LoRA-1/LoRA-2 track OPERA-CT/OPERA-GT closely), suggesting that training discovers two coherent end-to-end pipelines (audio expert plus generation adapter) rather than mixing components arbitrarily. Importantly, this does \emph{not} make the routers redundant, as they condition on different signals at different stages as reported in Section~\ref{sec:method_moa}

\input{tables/question_type_collapse}
\paragraph{Routing collapse analysis.}
\label{sec:collapse}
We further study \emph{routing collapse} disabling routing at inference by fixing an end-to-end path (audio expert $\in\{\text{OPERA-CT},\text{OPERA-GT}\}$, LoRA $\in\{\text{L1},\text{L2}\}$). \tableref{tab:question_types_collapse} reports Macro-F1 and MAE for forced routes (fixed audio expert $\times$ fixed LoRA). While some collapsed paths are locally competitive for certain tasks, they have a high variance across dataset-question-type combinations and the full model is strongest overall on regression and competitive on discriminative tasks. 
We also note that some fixed expert-adapter paths outperform the corresponding monolithic single-path baselines in \tableref{tab:main_results} on their strongest tasks (e.g. OPERA-GT for discriminative tasks), which suggests that specialization improves certain fixed-path configurations even before adaptive routing is applied. 

Our two-stage routing consistently yields the lowest MAE and remains competitive on discriminative tasks. By selectively leveraging expert-adapter pairs, the router balances performance across heterogeneous settings; for instance, it selects the high-performing OPERA-GT/L1 path for 88\% of KAUH samples. Crucially, sample-level routing can outperform even the best fixed path, as demonstrated by the superior regression results on the MM Lung dataset.

\subsection{Robustness and generalization}
\label{sec:generalization}
We assess robustness under controlled distribution shifts by comparing the strongest single-path baseline against our RAMoEA-QA model. 

\paragraph{Modality shift.}
\label{sec:modality_shift}
We evaluate \emph{unseen modality on a seen dataset and task} to probe robustness to changes of respiratory audio type within a dataset. Specifically, we focus on \textbf{Coswara asthma}, trained on breathing and tested here on cough, vowels and counting. While all models transfer reasonably, RAMoEA-QA remains competitive and more stable across runs. On \textbf{Coswara COPD} it improves Accuracy and Macro-F1 to $0.91$ and $0.92$ versus $0.68$ and $0.75$ (best baseline).

\paragraph{Dataset shift.}
\label{sec:dataset_shift}
We evaluate \emph{task seen during train but on an unseen dataset} to quantify transfer under cohort/device shift while keeping the question semantics fixed. Concretely, we test on \textbf{UK COVID-19} using asthma and COVID questions. RAMoEA-QA improves both tasks under this shift, reaching $0.88/0.88$ on asthma and $0.84/0.84$ on COVID, outperforming the strongest single-path baselines in both cases. This suggests that conditional specialization helps mitigate cohort/device differences when transferring to a new acquisition domain.

\paragraph{Task and modality shift.}
\label{sec:task_shift}
We evaluate \emph{an unseen task on unseen modalities} to measure zero-shot task transfer enabled by the shared generative interface. Concretely, we test \textbf{Coswara} on unseen symptom attributes (pneumonia),
and our model reaches the best performance of $0.83$ accuracy while one of the single-path baselines degrades to chance-level performance. 

\input{tables/generalization}

\subsection{Ablation studies} 
\paragraph{Comparison against a larger single adapter.}
To test whether the gains of RAMoEA-QA could be explained simply by increased adapter capacity, we compare it against a control with a \emph{single} LoRA adapter at \emph{double rank}, while keeping the rest of the architecture unchanged.  
As reported in Table~\ref{tab:appendix_controls}, increasing the rank of a single adapter improves performance slightly over the corresponding single-adapter setting ($0.68$ vs $0.70$ Accuracy, $0.86$ vs $0.88$ BERTScore), but does not recover the performance of our model. 
This suggests that the benefit of RAMoEA-QA does not arise from parameter increase alone, but from \emph{input-dependent expert selection} and the hierarchical specialization enabled by routing across both audio and language components.

\paragraph{Comparison against single-stage variants.}
We further evaluate whether the gains of RAMoEA-QA require \emph{both} routing stages, or whether specializing only the audio side or only the language side is sufficient. We compare the full two-stage model against two single-stage variants: \emph{MoA-only}, which uses a fixed audio encoder with two LoRA adapters, and \emph{MoE-only}, which uses the two audio experts independently with a single LoRA adapter. Results are reported in Table~\ref{tab:appendix_controls}. Both single-stage variants remain competitive but underperform the full model overall: the strongest MoA-only variant (\texttt{OPERA-GT + 2 LoRA}) reaches $0.70$ Accuracy, $2.58$ MAE, and $0.89$ BERTScore, while the MoE-only variant reaches $0.68$ Accuracy, $2.55$ MAE, and $0.86$ BERTScore. 
These results suggest that audio-side and language-side specialization are complementary, and that the strongest performance is obtained when both are combined within a unified two-stage routing framework.

\paragraph{Scaling experts and adapters.} \label{par:scaling}
 We further study how performance changes as we increase the number of audio experts and language adapters and report results in Table~\ref{tab:appendix_scaling}. Adding a third audio expert (\texttt{OPERA-CE}~\citep{zhang2024openrespiratoryacousticfoundation}) improves Accuracy and BERTScore ($0.72$ to $ 0.86$, $0.90 $ to $ 0.94$), but worsens MAE ($2.29$ to $2.55$), indicating a trade-off rather than a uniform gain. Increasing the number of adapters beyond two also does not yield consistent improvements: four adapters give similar Accuracy with slightly worse MAE, while eight adapters raise Accuracy to $0.75$ but further degrade MAE to $2.79$ without improving BERTScore beyond $0.91$. Overall, in our setting two adapters provide the best trade-off between specialization, stability, and performance.

\paragraph{Post-hoc ensemble control.}

To test whether the gains of RAMoEA-QA could be recovered by simply combining strong monolithic models after prediction, we implemented a lightweight post-hoc ensemble of the two strongest single-path baselines. The ensemble uses an agreement-plus-confidence tie-break rule: when both models produce the same answer, we return the shared prediction; otherwise, we select the prediction with the stronger available confidence proxy. As reported in Table~\ref{tab:appendix_ensemble}, this majority-vote baseline still underperforms RAMoEA-QA across both task families, reaching $0.61$ Accuracy, $0.54$ Macro-F1, and $2.62$ MAE, compared to $0.72$, $0.67$, and $2.29$ for RAMoEA-QA. These results suggest that the benefit of our approach does not come simply from merely combining multiple specialized models after prediction, but from \emph{input-dependent selection} of specialized sub-paths within a unified model.


%% file: tables/main_results.tex
\begin{table*}[t]
\tiny
\floatconts
  {tab:main_results}
  {\caption{\textbf{Main results.} We compare the single-path baselines and generic audio-language models to RAMoEA-QA, reporting both task- and text-level metrics. \textbf{NA} indicates that no valid numeric value could be parsed from the generated answer. \textbf{Bold} indicates the best result and \underline{underlining} indicates the second best. \textbf{*} indicates that the model was evaluated with an additional prompt. For generic audio-language models we report results for a single-run evaluation.
  }}

  {\begin{tabular}{lcc|cc|ccc}
\toprule
\bfseries Model & \multicolumn{4}{c|}{\bfseries Task-level metrics} & \multicolumn{3}{c}{\bfseries Text-level metrics} \\ \midrule
  & \multicolumn{2}{c|}{\bfseries Discriminative tasks} & \multicolumn{2}{c|}{\bfseries Regression tasks} & \multicolumn{3}{c}{\bfseries Global} \\
\midrule
 & Accuracy $\uparrow$ & Macro-F1 $\uparrow$ & MAE $\downarrow$ & RMSE $\downarrow$ & Token F1 $\uparrow$ & BERTScore $\uparrow$ & METEOR $\uparrow$ \\
\midrule
Pengi & 0.22 & 0.21 & NA & NA & 0.02 & -0.01 & 2.61 \\
LTU* & 0.18 & 0.11 & 14.36 & 36.20 & 0.34 & 0.90 & 52.64\\
GAMA & 0.45 & 0.28 & 2.83 & 4.21 & 0.43 & 0.76 & 70.63\\
\midrule
CareAQA-CT & 0.61 $\pm$ 0.01 & 0.53 $\pm$ 0.06 & 2.61 $\pm$ 0.16 & \underline{3.93 $\pm$ 0.14} & 0.83 $\pm$ 0.01 & 0.87 $\pm$ 0.01 & 84.89 $\pm$ 0.90 \\
CareAQA-GT & \underline{0.67 $\pm$ 0.15} & \underline{0.59 $\pm$ 0.20} & \underline{2.61 $\pm$ 0.09} & 4.26 $\pm$ 0.14 & \underline{0.86 $\pm$ 0.06} & \underline{0.89 $\pm$ 0.05} & \underline{87.05 $\pm$ 5.85} \\
\midrule
\textbf{RAMoEA-QA} & \textbf{0.72 $\pm$ 0.02} & \textbf{0.67 $\pm$ 0.03} & \textbf{2.29 $\pm$ 0.31} & \textbf{3.77 $\pm$ 0.53} & \textbf{0.88 $\pm$ 0.00} & \textbf{0.90 $\pm$ 0.01} & \textbf{88.38 $\pm$ 0.92} \\
\bottomrule
\end{tabular}}
\end{table*}

%% file: figures/figures_tex/tolerance.tex
\begin{figure*}[t]
\floatconts
  {fig:tolerance}
{\caption{\textbf{Tolerance accuracy for regression.} Accuracy as a function of absolute-error tolerance $\epsilon$ for spirometry targets (FVC/FEV1 on MM-Lung) and respiratory rate (NoseMic), comparing a single-path baseline against two-stage routing. Two-stage routing reaches higher accuracy at tighter tolerances, indicating fewer large prediction errors. }}
  {\includegraphics[width=\linewidth]{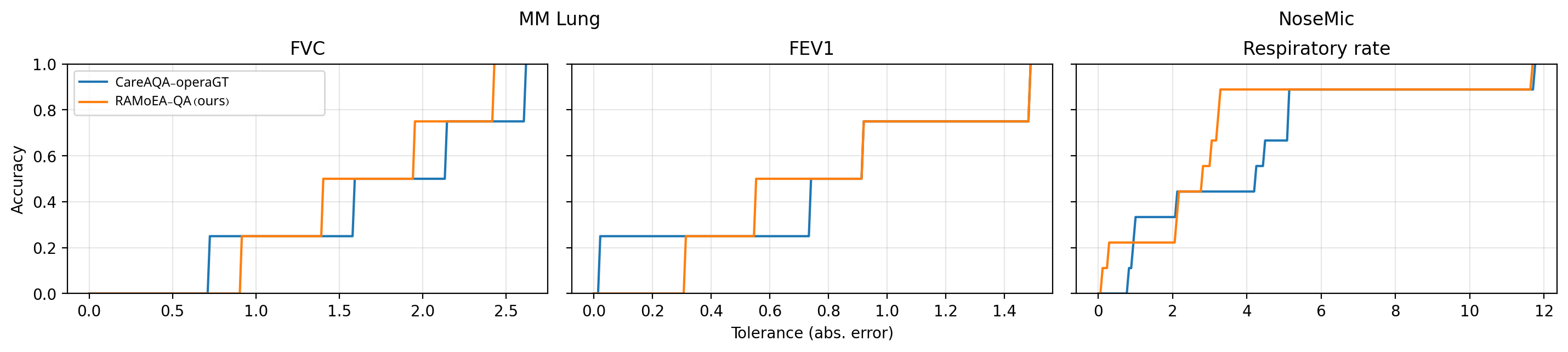}}
\end{figure*}

%% file: figures/figures_tex/general_routing.tex
\begin{figure*}[t]
\floatconts
  {fig:general_routing}
  {\caption{Unified routing heatmap across datasets, question formats, and diagnosis categories. Columns are grouped into Datasets (green), Question types (red), Diagnosis labels (blue), and Tasks (orange), while rows correspond to the four experts (operaCT, operaGT, LoRA expert 1, LoRA expert 2).}}
  {\includegraphics[width=0.9\linewidth]{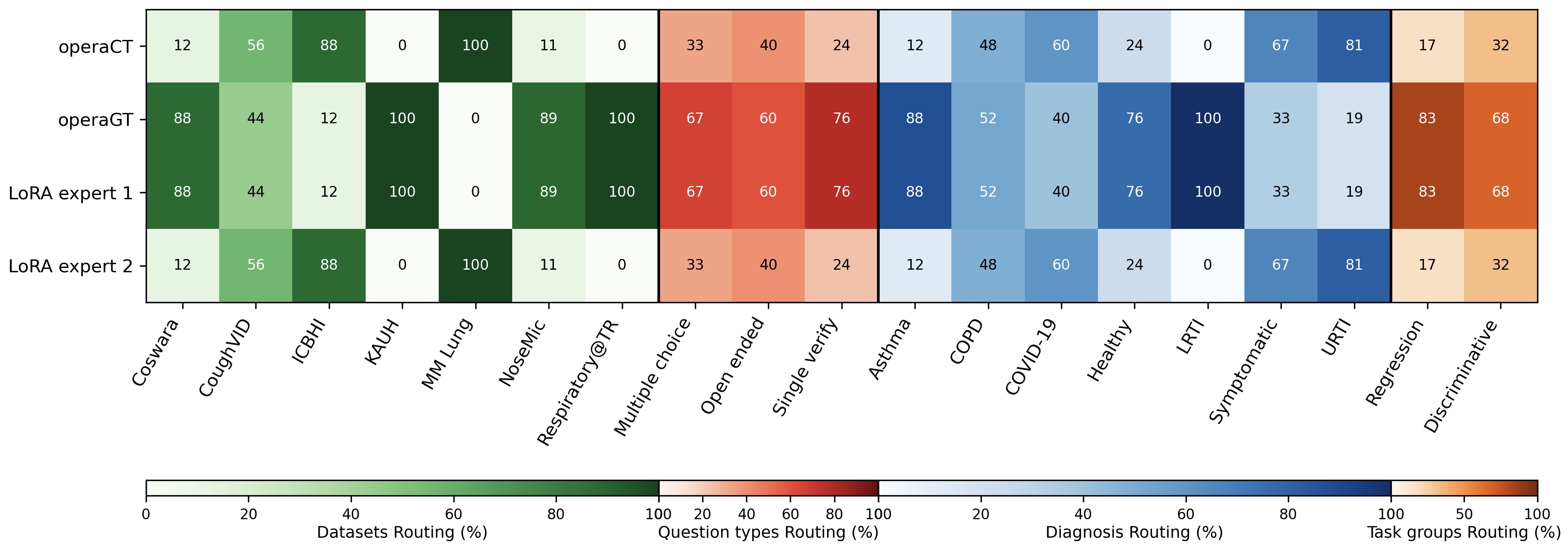}}
\end{figure*}

%% file: tables/question_type_collapse.tex
\begin{wraptable}{r}{0.48\textwidth}
\vspace{-0.9\baselineskip}
\centering
\tiny
\setlength{\tabcolsep}{2.5pt}
\renewcommand{\arraystretch}{0.92}

\caption{\textbf{Forced-route evaluation under expert collapse}. We report Macro-F1 (discriminative) and MAE (regression) by dataset and question type. Full is the fully routed model.}
\label{tab:question_types_collapse}

\begin{adjustbox}{width=\linewidth}
\begin{tabular}{llccccc}
\toprule
\bfseries Dataset & \bfseries QT & \multicolumn{4}{c}{\bfseries Forced routing} & \\ \midrule
 &  & \multicolumn{2}{c}{\bfseries operaCT} & \multicolumn{2}{c}{\bfseries operaGT} & \bfseries Full \\ \midrule
 & & L1 & L2 & L1 & L2 & \\ \midrule
\multirow{3}{*}{KAUH}     
  & SV & 0.75 & 0.75 & 1.00 & 1.00 & 1.00 \\
  & OE & 0.33 & 0.33 & 0.50 & 0.44 & 0.33 \\
  & MC & 0.33 & 0.33 & 0.44 & 0.50 & 0.38 \\ \midrule

\multirow{3}{*}{CoughVID} 
  & SV & 0.50 & 0.55 & 0.91 & 0.88 & 0.77 \\
  & OE & 0.16 & 0.16 & 0.16 & 0.00 & 0.00 \\
  & MC & 0.33 & 0.44 & 0.50 & 0.50 & 0.61 \\ \midrule

\multirow{3}{*}{ICBHI}    
  & SV & 0.75 & 0.75 & 1.00 & 1.00 & 0.71 \\
  & OE & 0.31 & 0.25 & 0.25 & 0.25 & 0.43 \\
  & MC & 0.50 & 0.50 & 0.50 & 0.50 & 0.43 \\ \midrule

\multirow{3}{*}{Resp.@TR} 
  & SV & 0.65 & 0.70 & 0.60 & 0.50 & 0.57 \\
  & OE & 1.00 & 1.00 & 1.00 & 1.00 & 1.00 \\
  & MC & 1.00 & 1.00 & 1.00 & 1.00 & 1.00 \\ \midrule

Coswara 
  & SV & 0.73 & 0.53 & 0.92 & 0.90 & 0.85 \\ \midrule

MM Lung 
  & OE ($\downarrow$) & 1.05 & 0.72 & 0.90 & 0.88 & 0.59 \\ \midrule

Nosemic 
  & OE ($\downarrow$) & 4.90 & 3.94 & 7.03 & 6.38 & 3.16 \\ \midrule

\multicolumn{2}{l}{\textbf{Disc. Avg.} (ALL)} 
  & 0.56 & 0.56 & 0.68 & 0.65 & 0.62 \\
\multicolumn{2}{l}{\textbf{Reg. Avg.} (OE $\downarrow$)} 
  & 2.98 & 2.33 & 3.97 & 3.63 & 1.88 \\
\bottomrule
\end{tabular}
\end{adjustbox}

\vspace{-0.8\baselineskip}
\end{wraptable}

%% file: tables/generalization.tex
\begin{table*}[t]
\centering
\fontsize{5.5pt}{6.1pt}\selectfont
\floatconts
  {tab:generalization}
 {\caption{\textbf{Robustness under controlled shifts.} We compare single-path baselines to RAMoEA-QA under modality, dataset, and task shifts. \textbf{Bold} indicates the best result and \underline{underlining} indicates the second best. \textbf{Legend:} $\Delta M$ = unseen modality; $\Delta D$ = unseen dataset; $\Delta TM$ = task and modalities never seen in training. }}
{\begin{tabular}{lllcc|cc|cc}
\toprule
\bfseries Shift &\bfseries Dataset  &\bfseries Task    & \multicolumn{2}{c}{CareAQA-operaCT}   & \multicolumn{2}{c}{CareAQA-operaGT}   &\multicolumn{2}{c}{\bfseries RAMoEA-QA (ours)}\\ \midrule \addlinespace[0.2em]
   &   &   &  Accuracy  $\uparrow$ &  MacroF1  $\uparrow$ &  Accuracy  $\uparrow$ & MacroF1  $\uparrow$ &  Accuracy  $\uparrow$   & MacroF1  $\uparrow$ \\ \midrule 
  \addlinespace[0.2em]
\multirow{2}{*}{$\Delta M$} & \multirow{2}{*}{Coswara}  & Asthma& \textbf{0.90 ± 0.13}   & \textbf{0.90 ± 0.13} & 0.66 ± 0.47 & 0.75 ± 0.35 & \underline{0.87 ± 0.02}   & \underline{0.87 ± 0.02} \\ \addlinespace[0.2em]
  &   & COPD  & \underline{0.68 ± 0.44}   & \underline{0.75 ± 0.34} & 0.66 ± 0.47 & 0.75 ± 0.35 & \textbf{0.91 ± 0.02  }  & \textbf{0.92 ± 0.02} \\ \midrule \addlinespace[0.2em]
\multirow{2}{*}{$\Delta D$}  & \multirow{2}{*}{UK COVID-19} & Asthma& \underline{0.84 ± 0.21} & \underline{0.84 ± 0.21} & 0.66 ± 0.47 & 0.74 ± 0.35 & \textbf{0.88 ± 0.07} & \textbf{0.88 ± 0.06} \\ \addlinespace[0.2em]
  &   & COVID & 0.56 ± 0.32 & 0.65 ± 0.20 & \underline{0.66 ± 0.47} & \underline{0.74 ± 0.35} & \textbf{0.84 ± 0.04} & \textbf{0.84 ± 0.04} \\ \midrule \addlinespace[0.2em]
\multirow{1}{*}{$\Delta TM$} & Coswara  & Pneumonia & \underline{0.77 ± 0.31} & 0.79 ± 0.28 & 0.33 ± 0.00 & 0.50 ± 0.00 & \textbf{0.83 ± 0.05}  & \textbf{0.84 ± 0.04} \\ \midrule
\multicolumn{1}{l}{\textbf{Avg.}} & & & \underline{0.75 ± 0.19} & \underline{0.75 ± 0.19}  & 0.67 ± 0.20 & 0.59 ± 0.29 &   \textbf{0.78 ± 0.04} & \textbf{0.79 ± 0.04} \\ 
\bottomrule
\end{tabular}}
\end{table*}

%% file: sections/6_conclusions.tex
\section{Discussion} \label{sec:discussion}
We presented \textbf{RAMoEA-QA}, a unified framework for respiratory audio question answering that jointly adapts acoustic processing and language generation to heterogeneous recordings and query types. We showed that this hierarchical two-stage specialization improves performance over both generic audio-language models and matched monolithic single-path baselines, while retaining efficient single-path inference. 

These gains hold not only in-distribution, but also under controlled dataset, modality, and task shifts, showing that RAMoEA-QA transfers more reliably across heterogeneous respiratory audio settings without modifying the frozen backbones. Across both \emph{discriminative} and \emph{regression} question families, the results indicate that input-dependent specialization provides a robust and effective design for respiratory audio QA.

Future directions include exploiting \emph{router uncertainty} as a confidence signal for \emph{selective answering} (abstention) and studying how the proposed architecture scales with larger backbones, more experts, and additional data.

%% file: appendix/datasets_overview.tex
\section{Datasets overview}\label{apd:datasets_overview}
The RA-QA collection is built upon 11 distinct datasets, each contributing to the overall diversity and richness of the resource (dataset availability is reported in \tableref{tab:datasets_availability}). \input{tables/appendix_supplementary/datasets_availability}
This section provides a more detailed overview of these datasets. Each dataset comprises two core components: a metadata file and a corresponding audio set. The metadata encompasses structured clinical and contextual information, while the audio data consists of respiratory-related recordings. All audio recordings are fully anonymized and the associated metadata have been carefully curated to exclude any personally identifiable information or inappropriate content, ensuring compliance with ethical standards and data protection regulations. This unified structure supports consistent integration and facilitates downstream research tasks. Following this introduction, a more detailed description is provided for each dataset from the RA-QA collection used in this paper. \tableref{tab:data_stats} reports the number of recordings and QA pairs used per dataset, split, modality, and task family within training.\\
\input{tables/appendix_supplementary/dataset_statistics}

\textbf{UK COVID-19} \citep{coppock2023audiobasedaiclassifiersevidence}. The UK COVID-19 Vocal Audio Dataset is designed for training and evaluating machine learning models to classify SARS-CoV-2 infection status and associated respiratory symptoms using vocal audio. The dataset was collected by the UK Health Security Agency from March 2021 to March 2022, during the prevalence of the Alpha and Delta SARS-CoV-2 variants, as well as some Omicron sublineages. Participants were recruited through the national Test and Trace programme and the REACT-1 survey. Audio recordings of voluntary coughs and exhalations were gathered through the 'Speak up to help beat coronavirus' digital survey, alongside demographic data, self-reported symptoms and respiratory conditions (see \figureref{fig:uk_covid_19_summary_plot}). These recordings were linked to SARS-CoV-2 test results, although speech recordings are not included in the open access version of the dataset and were not used in this study. The study was approved by The National Statistician's Data Ethics Advisory Committee (reference NSDEC(21)01), the Cambridge South NHS Research Ethics Committee (reference 21/EE/0036) and the Nottingham NHS Research Ethics Committee (reference 21/EM/0067). Informed consent was obtained from all participants. 
\\ \\
\textbf{COUGHVID} \citep{orlandic_coughvid_2021} The COUGHVID dataset consists of over 25,000 crowdsourced cough recordings, collected from a diverse group of participants across different ages, genders, geographic locations, and COVID-19 statuses (see \figureref{fig:coughvid_summary_plot}). All data collection and annotation processes adhered to the relevant ethical guidelines, with informed consent obtained from all participants who submitted their cough recordings and associated metadata.
\\ \\
 \textbf{ICBHI} \citep{rocha_open_2019} The ICBHI Respiratory Sound Database includes audio samples collected by two independent research teams from different countries over several years. Ethical approval was granted by the relevant ethics committees of the respective institutions. The majority of the database comprises audio samples recorded by the School of Health Sciences, University of Aveiro (ESSUA) research team at the Respiratory Research and Rehabilitation Laboratory (Lab3R) in Aveiro, Portugal and Hospital Infante D. Pedro. The second team, from Aristotle University of Thessaloniki (AUTH) and the University of Coimbra (UC), collected respiratory sounds at Papanikolaou General Hospital in Thessaloniki and the General Hospital of Imathia (Health Unit of Naousa), Greece. The database contains a total of 5.5 hours of recordings across 920 annotated audio samples from 126 subjects (see \figureref{fig:icbhi_summary_plot}).
\\ \\
\textbf{MM Lung} \citep{mosuily23_interspeech}: The MMLung dataset was collected from 40 participants (20 male, 20 female) aged 18 to 85 years, all of whom are English speakers from the UK. Among the participants, 12 were healthy, while the others included seven self-reported COPD patients, seven self-reported asthma patients, and 14 individuals with other long-term conditions. Ethical approval for this study was granted by the University of Southampton. Data were gathered using three devices: a Google Pixel 6 smartphone with a custom app for data collection and an Easy on-PC ultrasonic spirometer from ndd Medical Technologies. The audio data was collected in stereo mode at a sampling rate of 44,100 Hz, saved in WAV format and recorded in a quiet room to ensure optimal conditions. The data collection included four audio modalities: cough, vowels, mobile spirometry and speech, through a series of tasks performed in a single session by each participant. For this paper, only the deep breath and the vowel sound of 'o' are included  (see \figureref{fig:mm_lung_summary_plot}). Ground truth data were obtained using a medical-grade spirometer, with measurements taken by a healthcare professional according to European Respiratory Society (ATS/ERS) clinical standards. However, it is important to note that objective measurements can be subject to individual effort, which may introduce some errors (e.g., effort-dependent blows). This dataset is available upon request. 
\\ \\
\textbf{Coswara} \citep{bhattacharya_coswara_2023} The Coswara dataset consists of respiratory sounds recorded between April 2020 and February 2022 from 2,635 individuals, including 1,819 SARS-CoV-2 negative, 674 positive and 142 recovered participants. The dataset includes nine categories of respiratory sounds related to breathing, coughing, and speech. Metadata accompanying the recordings contains demographic details such as age, gender and geographic location, as well as health-related information including symptoms, pre-existing respiratory conditions, comorbidities and SARS-CoV-2 test status  (see \figureref{fig:coswara_summary_plot}). The data collection was approved by the Institutional Human Ethics Committee at the Indian Institute of Science, Bangalore. Informed consent was obtained from all participants who uploaded their data and the collected data was anonymized to exclude personally identifiable information.
\\ \\
\textbf{KAUH} \citep{fraiwan_dataset_2021} The KAUH dataset includes audio recordings of lung sounds from patients with seven different respiratory conditions: asthma, heart failure, pneumonia, bronchitis, pleural effusion, lung fibrosis and chronic obstructive pulmonary disease (COPD), as well as normal breathing sounds (see \figureref{fig:KAUH_summary_plot}). The recordings were made using an electronic stethoscope, with the chest wall examined at various points by a specialist physician. Each sound was recorded three times with different frequency filters to highlight specific bodily sounds. This dataset is valuable for the development of automated systems designed to detect pulmonary diseases or classify lung sounds. All participants (or their parents in the case of minors), provided written informed consent for their inclusion in the study and the sharing of their data. The study was approved by the Institutional Review Board at King Abdullah University Hospital and Jordan University of Science and Technology (Ref. 91/136/2020). Data collection adhered to all relevant ethical guidelines and regulations and the authors have the right to share the data publicly.
\\ \\
\textbf{NoseMic} \citep{butkow_evaluation_2024} The NoseMic dataset is a subset of data collected for a respiratory rate estimation project. The audio recordings were captured using microphones positioned close to the participants' noses, while respiratory dynamics were measured using a Zephyr pressure sensor on the chest. Data collection took place in a stationary environment, both before and after physical exercise. A total of 21 participants were involved in the study, with data from some participants excluded due to poor sensing quality. The benchmark includes audio recordings from before and after running, with each recording segmented into 30-second windows that overlap by 15 seconds. The average respiratory rate for each window was used as the ground truth.
\\ \\
\textbf{Respiratory@TR} \citep{altan2021multimediarespiratorydatabaserespiratorydatabasetr}
The Respiratory@TR dataset contains lung sounds recorded from both the left and right sides of the posterior and anterior chest wall, as well as the back, using two digital stethoscopes at Antakya State Hospital. In addition to the lung sound recordings, the dataset includes chest X-rays, pulmonary function test (PFT) variables, spirometric curves and the St. George Respiratory Questionnaire (SGRQ-C) as multimedia and clinical functional analysis data. The lung sounds were captured across 12 channels, focusing on the upper, middle and lower lung regions, as well as the costophrenic angles on both the posterior and anterior chest sides. The recordings were validated and labeled by two pulmonologists, who assessed the chest X-rays, PFT results and auscultation sounds of the subjects. The labels correspond to five levels of COPD severity (COPD0, COPD1, COPD2, COPD3, COPD4). The dataset was released by Iskenderun Technical University, Turkey. Participation was voluntary and patients, aged 38 to 68, were selected from various occupational groups, socio-economic backgrounds and genders to ensure a diverse representation of the disorders.

\input{figures/figures_tex/appendix_datasets/uk_covid_19}
\input{figures/figures_tex/appendix_datasets/coughvid}
\input{figures/figures_tex/appendix_datasets/icbhi}
\input{figures/figures_tex/appendix_datasets/mm_lung}
\input{figures/figures_tex/appendix_datasets/coswara}
\input{figures/figures_tex/appendix_datasets/kauh}
\input{figures/figures_tex/appendix_datasets/nosemic}
\input{figures/figures_tex/appendix_datasets/respiratoryTR}

%% file: tables/appendix_supplementary/datasets_availability.tex
\begin{table*}[htbp]
\scriptsize
\floatconts
  {tab:datasets_availability}%
  {\caption{Dataset Availability: The ICBHI and HF Lung datasets are sourced from multiple origins; please refer to the detailed description in the text below. The MMLung and NoseMic datasets are available upon request. The specific licensing terms are outlined in the Data Transfer Agreement (DTA).
  }}%
  {%
  \begin{tabular}{llll}
  \toprule
  \bfseries Dataset & \bfseries Source & \bfseries Access & \bfseries License\\
  \midrule
UK COVID-19 & IC & \url{https://zenodo.org/records/10043978} & OGL 3.0 \\
CoughVID & EPFL & \url{https://zenodo.org/records/4048312} & CC BY 4.0 \\
ICBHI & * & \url{https://bhichallenge.med.auth.gr} & CC0 \\
\multirow{2}{*}{HF Lung} & \multirow{2}{*}{*} 
    & \url{https://gitlab.com/techsupportHF/HF_Lung_V1} & CC BY 4.0 \\
  &  & \url{https://gitlab.com/techsupportHF/HF_Lung_V1_IP} & CC BY-NC 4.0 \\
Coswara & IISc & \url{https://github.com/iiscleap/Coswara-Data} & CC BY 4.0 \\
KAUH & KAUH & \url{https://data.mendeley.com/datasets/jwyy9np4gv/3} & CC BY 4.0 \\
Respiratory@TR & ITU & \url{https://data.mendeley.com/datasets/p9z4h98s6j/1} & CC BY 4.0 \\
MMlung & UoS & \url{https://github.com/MohammedMosuily/mmlung} & Custom license \\
NoseMic & UoC & \url{https://github.com/evelyn0414/OPERA/tree/main/datasets/nosemic} & Custom license \\
  \bottomrule
  \end{tabular}
  }
\end{table*}

%% file: tables/appendix_supplementary/dataset_statistics.tex
\begin{table*}[hbtp]
\tiny
\floatconts
  {tab:data_stats}
  {\caption{Statistics of the RA-QA collection used in the training. For each dataset we report the task family, target task, audio modalities, and the number of derived QA pairs per split (Train/Val/Test). An asterisk (*) denotes datasets where the diagnosis task is available only in the single-verify format.} }
\centering
  {\begin{tabular}{lllll}
  \toprule
   \bfseries Task   & \bfseries Dataset & \bfseries Task   & \bfseries Modalities & \bfseries QA $N$ \\  \midrule
\multirow{5}{*}{Discr.} & Coswara   & Asthma * & Deep and shallow breathing & 204 / 64 / 64\\
& COUGHVID  & Diagnosis (Healthy, Symptomatic, Covid-19) & Coughing & 225 / 45 / 45     \\
& ICBHI     & Diagnosis (URTI, COPD) & Four clinical recording devices & 423 / 32 / 32     \\
& KAUH& Diagnosis (Healthy, LRTI, COPD) & 6 sound types and 7 chest positions & 225 / 45 / 45     \\
& Respiratory@TR  & COPD Severity  & 11 recording positions & 210 / 70 / 70     \\ \midrule
\multirow{3}{*}{Regr.}     & MM-Lung   & Spirometry (FEV1)   & Deep breathing & 32 / 4 / 4  \\
& MM-Lung   & Spirometry (FVC)    & Deep breathing & 32 / 4 / 4  \\
& Nosemic   & Respiratory rate & Before and after running  & 50 / 9 / 9  \\ \bottomrule
\end{tabular}}
\end{table*}

%% file: figures/figures_tex/appendix_datasets/uk_covid_19.tex
\begin{figure*}[htbp]
\floatconts
  {fig:uk_covid_19_summary_plot}
  {\caption{The figure shows label distributions for viral load categories (multiclass), as well as binary labels for symptoms such as runny or blocked nose and conditions like asthma from the UK COVID-19 dataset. As with other datasets used in this work, these labels are highly imbalanced and require preprocessing and reduction strategies to ensure meaningful training and evaluation.
  }}
  {\includegraphics[width=1\linewidth]{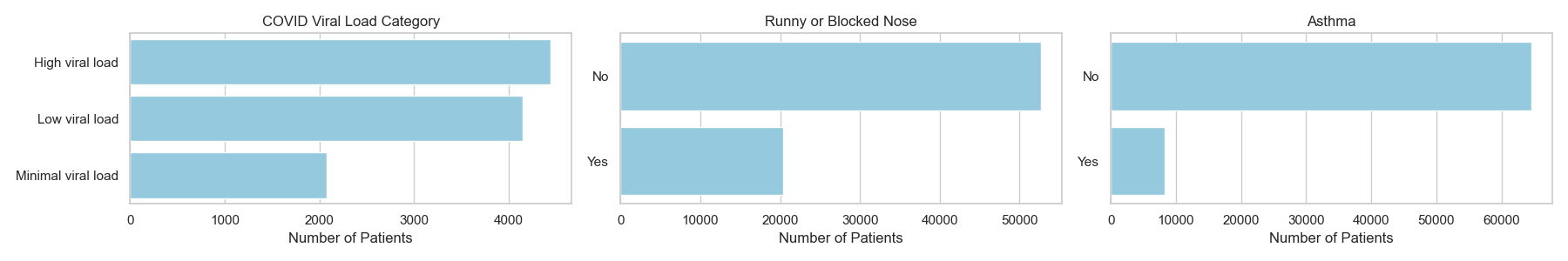}}
\end{figure*}

%% file: figures/figures_tex/appendix_datasets/coughvid.tex
\begin{figure*}[htbp]
\floatconts
  {fig:coughvid_summary_plot}
  {\caption{Label distributions from the CoughVid dataset, illustrating examples of audio-related attributes such as cough type, presence of stridor and associated diagnoses. This highlights that, beyond clinical metadata, some datasets also include perceptual or acoustic labels (e.g., wheezes, stridors), which are directly linked to the audio signal and can support more fine-grained sound analysis.
  }}
  {\includegraphics[width=1\linewidth]{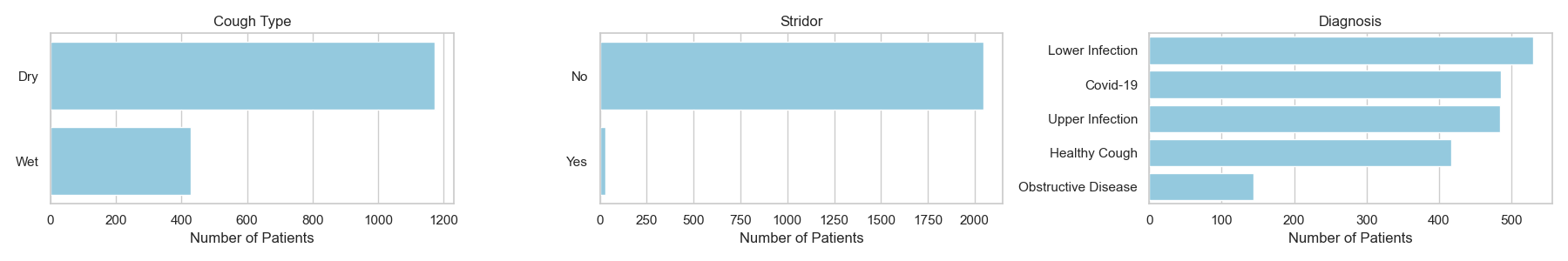}}
\end{figure*}

%% file: figures/figures_tex/appendix_datasets/icbhi.tex
\begin{figure*}[htbp]
\floatconts
  {fig:icbhi_summary_plot}
  {\caption{Label distributions from the ICBHI dataset, showing the annotated diagnosis categories, presence of crackles and recording positions (e.g., trachea, anterior left, posterior right). This exemplifies how some datasets provide detailed contextual metadata, such as auscultation position, which can be crucial for interpreting respiratory sounds and modeling location-sensitive acoustic features.
  }}
  {\includegraphics[width=1\linewidth]{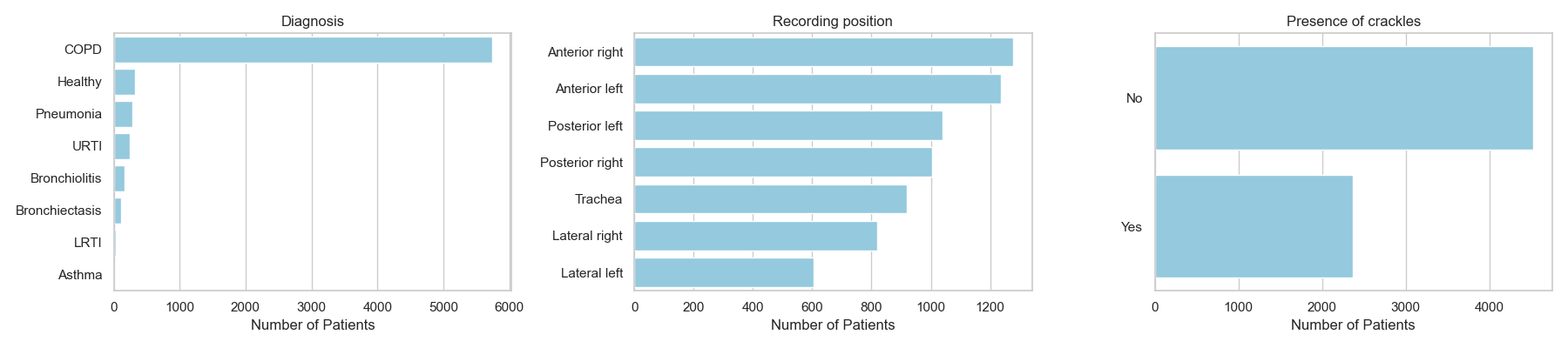}}
\end{figure*}

%% file: figures/figures_tex/appendix_datasets/mm_lung.tex
\begin{figure*}[htbp]
\floatconts
  {fig:mm_lung_summary_plot}
  {\caption{The figure illustrates key characteristics from the MM Lung dataset.The Type of Audio distinguishes between different types of respiratory recordings, such as vowel phonation and deep breathing, which capture varied aspects of lung sound dynamics.
  }}
  {\includegraphics[width=1\linewidth]{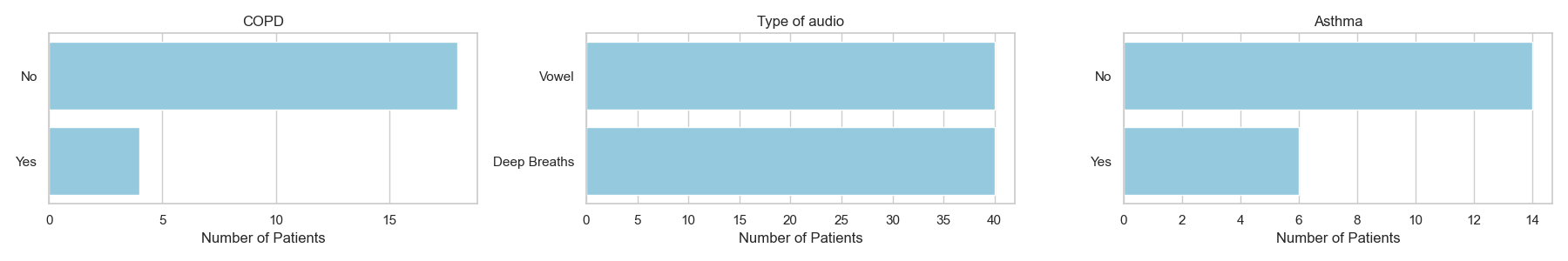}}
\end{figure*}

%% file: figures/figures_tex/appendix_datasets/coswara.tex
\begin{figure*}[htbp]
\floatconts
  {fig:coswara_summary_plot}
  {\caption{The figure presents the distribution from the Coswara dataset of the selected attributes health status, COVID test result and mask usage at the time of recording. The Health Status attribute distinguishes between healthy individuals and COVID-19 positive cases with varying symptom severity, categorized as asymptomatic, mild or moderate.
  }}
  {\includegraphics[width=1\linewidth]{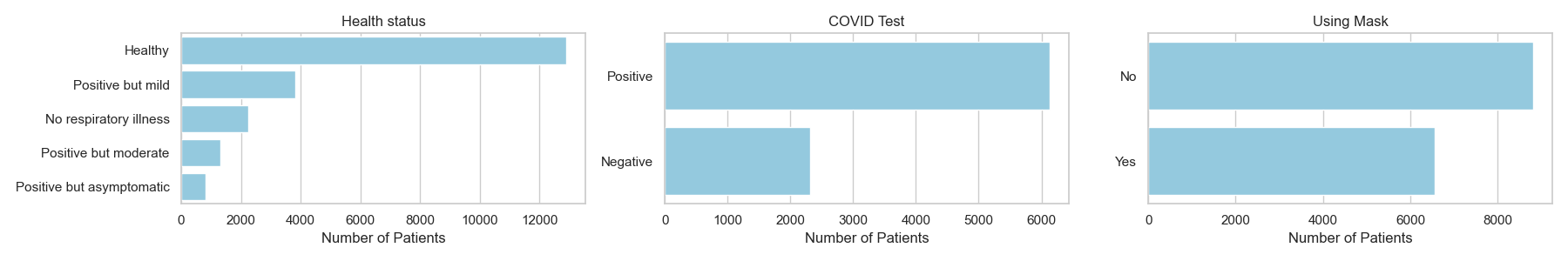}}
\end{figure*}

%% file: figures/figures_tex/appendix_datasets/kauh.tex
\begin{figure*}[htbp]
\floatconts
  {fig:KAUH_summary_plot}
  {\caption{The figure displays key attributes of the KAUH dataset: sound type, recording position and diagnosis. The Sound Type plot shows the presence of multiple annotated labels per recording.
  }}
  {\includegraphics[width=1\linewidth]{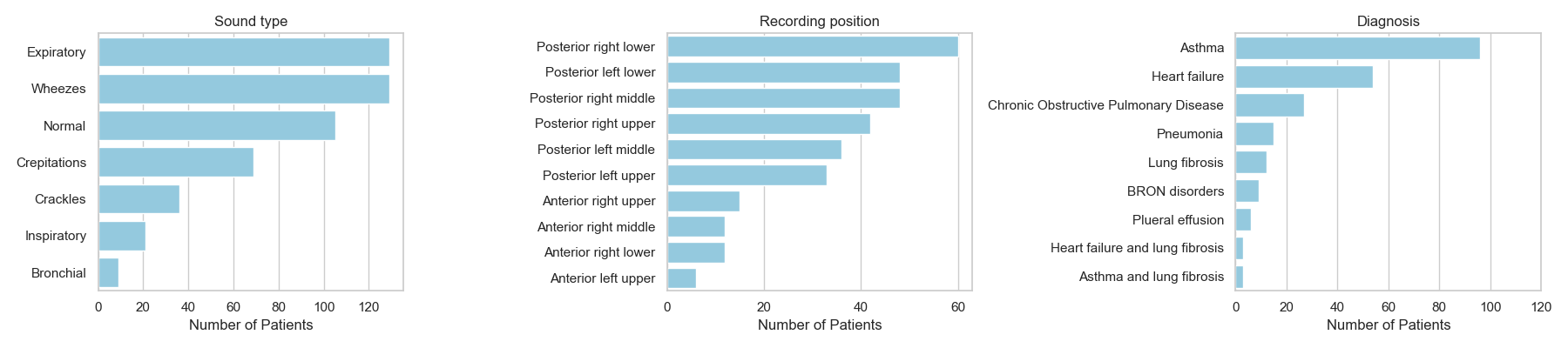}}
\end{figure*}

%% file: figures/figures_tex/appendix_datasets/nosemic.tex
\begin{figure*}[htbp]
\floatconts
  {fig:nosemic_summary_plot}
  {\caption{The plots illustrate the recording phase attribute distribution form the NoseMic dataset. This captures whether the respiratory sounds were recorded before or after physical exertion (e.g., running), which is important for evaluating exertion-induced changes in breathing patterns.
  }}
  {\includegraphics[width=0.3\linewidth]{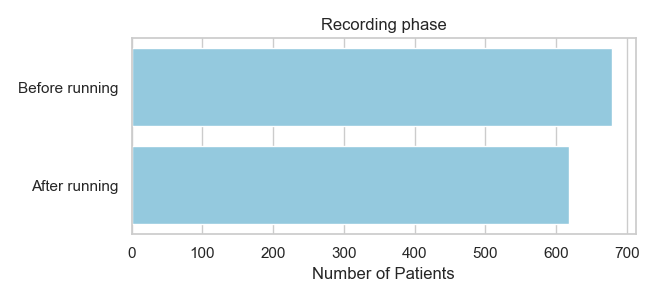}}
\end{figure*}

%% file: figures/figures_tex/appendix_datasets/respiratoryTR.tex
\begin{figure*}[htbp]
\floatconts
  {fig:respiratoryTR_summary_plot}
  {\caption{The plots illustrate two key attributes from the Respiratory@TR dataset: recording position and COPD severity. The recording position captures the recording point while the COPD Severity is scored on a scale from 0 (no COPD) to 4 (very severe), providing clinically relevant gradation to analyze how disease progression correlates with acoustic features in the recordings.
  }}
  {\includegraphics[width=0.6\linewidth]{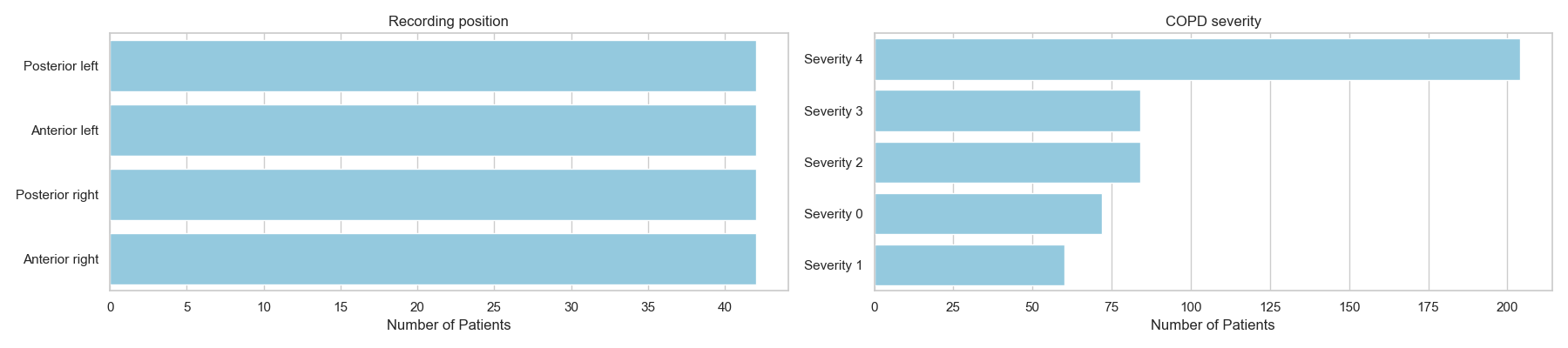}}
\end{figure*}

%% file: appendix/metrics_overview.tex
\section{Metrics overview} \label{apd:metrics_overview}
We follow the RA-QA evaluation protocol \citep{Bertolino_Zang_Xia_Talia_Mascolo_2026} and report both \emph{text-level} and \emph{task-level} metrics to separately capture (i) how well the model matches the reference form and semantics, and (ii) whether it predicts the correct clinical label or value. Text-level metrics are computed directly between the generated answer and the reference after standard normalization, while task-level metrics operate on an \emph{extracted} label/value from the output and therefore ignore any additional explanatory context.

\subsection{Metrics for discriminative tasks.}
We group discriminative evaluation into \emph{response form} metrics and \emph{label-correctness} metrics.

\subsubsection{Response form metrics.}
These metrics assess surface fidelity and semantic similarity of the generated answer:
\begin{itemize}
    \item \textbf{Exact Match (EM)} evaluates whether the normalized predicted response matches the normalized reference string exactly (strict match). EM is informative for structured outputs but can underestimate performance for valid paraphrases.
    \item \textbf{TokenF1} measures token-level overlap between prediction and reference using the harmonic mean of token precision and token recall. TokenF1 rewards partial matches and is less brittle than EM.
    \item \textbf{METEOR} measures lexical alignment between prediction and reference, designed to be robust to minor morphological variation and alternate word forms via flexible word-level matching.
    \item \textbf{BERTScore} \citep{zhang2020bertscoreevaluatingtextgeneration} measures semantic similarity by computing cosine similarity between contextual token embeddings from a pre-trained BERT model and aggregating alignment scores across tokens. It provides a semantic view that is less sensitive to paraphrasing than n-gram overlap metrics.
\end{itemize}
Together, EM/TokenF1 emphasize surface-form fidelity, METEOR captures word-level matching under small lexical variations, and BERTScore provides a semantics-oriented complement for open-ended answers.

\subsubsection{Label-correctness metrics.}
For diagnosis-style questions, we evaluate \emph{label correctness} by extracting the target label from the model output and comparing it to the ground truth, ignoring extra answer context. For open-ended and multiple-choice questions, we parse the generated text to recover the predicted diagnosis label (e.g., by normalizing case/punctuation and matching against the dataset label set), whereas for single-verify questions we parse the output into $\{\texttt{Yes},\texttt{No}\}$. We then compute:
\begin{itemize}
    \item \textbf{Accuracy} as the fraction of examples whose extracted label matches the ground truth. While interpretable, accuracy can be misleading under class imbalance.
    \item \textbf{Macro F1-score (MacroF1)} as the unweighted mean of per-class F1-scores, where each class contributes equally. MacroF1 is robust to imbalance and reflects the precision--recall trade-off, which is important in clinical settings where both missed cases (false negatives) and false alarms (false positives) carry cost.
\end{itemize}
We emphasize \textbf{MacroF1} as the primary discriminative metric due to its robustness to skewed label distributions and its balanced accounting of false positives and false negatives.

\subsection{Metrics for regression tasks.}
For numeric targets, we treat the model output as free-form text and \emph{parse} a scalar prediction by extracting the first valid numeric value in the generated answer (if present). We then compare this extracted value $\hat{y}$ to the ground-truth target $y$ and compute error metrics over successfully parsed examples (lower is better). Concretely, we report:
\begin{itemize}
    \item \textbf{Mean Absolute Error (MAE)}: $\mathrm{MAE}=\frac{1}{N}\sum_{i=1}^{N}|\hat{y}_i-y_i|$, which measures the average absolute deviation and is less sensitive to occasional outliers.
    \item \textbf{Root Mean Squared Error (RMSE)}: $\mathrm{RMSE}=\sqrt{\frac{1}{N}\sum_{i=1}^{N}(\hat{y}_i-y_i)^2}$, which penalizes larger errors more strongly and is therefore sensitive to rare but severe mis-estimates.
\end{itemize}
Because regression evaluation depends on successful numeric extraction from free-form generations, we additionally report \textbf{parsing coverage}, i.e., the fraction of test instances for which numeric extraction succeeds. Unless otherwise stated, MAE/RMSE are computed on the subset of examples with valid numeric parses.

%% file: appendix/method.tex
\section{Routing and conditioning details}
\label{apd:method_appendix_math}

\paragraph{Audio-MoE routing vector.}
We denote the Audio-MoE routing vector by $\mathbf{r}_a\in\mathbb{R}^{d_r}$ and instantiate the router-input policy as:
\begin{equation}
\mathbf{r}_a \in \{\mathbf{z}^{a}_{\text{audio}},\ \mathbf{z}^{a}_{\text{question}},\ \mathbf{z}^{a}_{\text{fused}}\}\,,
\end{equation}
where $\mathbf{z}^{a}_{\text{audio}}$ is obtained from a shallow spectrogram proxy extractor $\mathcal{F}$ and pooling, $\mathbf{z}^{a}_{\text{question}}$ is a pooled question embedding projected to $d_r$, and $\mathbf{z}^{a}_{\text{fused}}$ is computed via lightweight cross-attention between question tokens and proxy audio tokens. The router MLP produces logits over $N_a$ experts and selects a single expert via straight-through Gumbel-Softmax in training and argmax at inference.

\paragraph{Audio expert encoding and alignment.}
Given the selected expert index $e_a$, only $\mathcal{E}^{a}_{e_a}$ encodes the spectrogram to produce $\mathbf{h}^{a}_{e_a}$, which is aligned to the LLM hidden size $d_\ell$ with an expert-specific aligner $\mathcal{A}_{e_a}$:
\begin{equation}
\tilde{\mathbf{h}}^{a}_{e_a} = \mathcal{A}_{e_a}(\mathbf{h}^{a}_{e_a}) \in \mathbb{R}^{T_{e_a}\times d_\ell}.
\end{equation}

\paragraph{Selected audio prefix and LLM input.}
We inject the aligned audio embeddings as a \emph{selected audio prefix} by concatenating them into the LLM input embeddings:
\begin{equation}
\mathbf{X} = [\mathbf{H}_p;\ \tilde{\mathbf{h}}^{a}_{e_a};\ \mathbf{H}_{\texttt{Ans}};\ \mathbf{H}_y],
\end{equation}
where $\mathbf{H}_p$ encodes the prompt/question, $\mathbf{H}_{\texttt{Ans}}$ is an ``Answer:'' tag, and $\mathbf{H}_y$ are teacher-forced answer embeddings during training.

\paragraph{LoRA-MoA routing vector.}
Analogously, the MoA routing vector $\mathbf{r}_\ell\in\mathbb{R}^{d_\ell}$ instantiates the same router-input policy at the language stage:
\begin{equation}
\mathbf{r}_\ell \in \{\mathbf{z}^{\ell}_{\text{audio}},\ \mathbf{z}^{\ell}_{\text{question}},\ \mathbf{z}^{\ell}_{\text{fused}}\},
\end{equation}
with $\mathbf{z}^{\ell}_{\text{audio}}=\mathrm{Pool}(\tilde{\mathbf{h}}^{a}_{e_a})$, $\mathbf{z}^{\ell}_{\text{question}}=\mathrm{Pool}(\mathbf{H}_p)$, and $\mathbf{z}^{\ell}_{\text{fused}}$ computed by lightweight fusion between prompt and the selected aligned audio embeddings. The adapter router outputs logits over $N_\ell$ LoRA adapters and selects one adapter via straight-through Gumbel-Softmax (train) and argmax (eval).

%% file: appendix/baselines.tex
\section{Baselines}
\label{apd:baselines}

We compare RAMoEA-QA against three representative baselines: (i) a general-purpose audio-language model used \emph{as-is} (\textsc{Pengi}), (ii) a stronger general-purpose large audio-language model with instruction tuning and complex audio reasoning capabilities (\textsc{GAMA}), and (iii) a domain-specific audio-QA architecture implemented as a \emph{monolithic} (single-path) counterpart to our routed model (\textsc{CaReAQA-style}).

\subsection{\textsc{Pengi} (general-purpose audio-language model)}
\label{apd:baseline_pengi}
\textsc{Pengi} is a general audio-language model originally designed for broad audio captioning and QA-style generation across diverse audio domains (see \figureref{fig:pengi}). It formulates a wide range of audio tasks as \emph{audio+text $\rightarrow$ text} generation by conditioning a (largely) frozen causal language model on a fixed-length \emph{prefix} built from an audio embedding and a text-prompt embedding, each mapped through lightweight trainable networks \citep{deshmukh_pengi_2024}. In our experiments, we use \textsc{Pengi} \emph{without additional fine-tuning} on RA-QA: we provide the respiratory recording and question as input and decode answers using the model's default generation procedures. Because \textsc{Pengi} is not trained for clinically structured respiratory assessment (nor specialized to RA-QA labels, question formats, or respiratory acoustics), it serves as a reference point for the limitations of general multimodal audio-language models when transferred to respiratory clinical QA. This baseline therefore tests how far a strong, task-general audio-language model transfers under dataset, modality, and task shifts.
\input{figures/figures_tex/appendix_supplementary/pengi}

\subsection{\textsc{GAMA} (general-purpose large audio-language model)}
\label{apd:baseline_gama}
\textsc{GAMA} is a general-purpose large audio-language model designed for advanced audio understanding and complex reasoning \citep{ghosh2024gama}. Its architecture integrates a large language model with multiple audio representations, including features extracted through a custom Audio Q-Former and a multi-layer aggregation mechanism over an audio encoder, and is further instruction-tuned to improve audio question answering and reasoning over complex auditory scenes \citep{ghosh2024gama}. In our experiments, we use \textsc{GAMA} \emph{without additional fine-tuning} on RA-QA: we provide the respiratory recording and question as input and decode answers using the model's default generation procedures. Because \textsc{GAMA} is designed as a broad audio-language model rather than a clinically specialized respiratory QA system, it serves as a stronger general-purpose baseline than \textsc{Pengi}, testing whether improved generic audio reasoning transfers to structured respiratory assessment across datasets, modalities, and tasks.

\subsection{\textsc{CaReAQA-style} monolithic audio-QA model}
\label{apd:baseline_careaqa}
\textsc{CaReAQA} is an audio-language QA framework for cardiac and respiratory diagnostic reasoning that combines an audio encoder with a decoder-only LLM to produce open-ended answers \citep{Wang_Chen_Zeghidour_Saeed_2025}. Its core design maps audio features into the LLM hidden space via an audio mapper and injects the resulting aligned embeddings as a soft prefix for autoregressive generation.
Prior work in this line is typically trained and evaluated primarily in a monolithic setting and on a single question format (open-ended), and does not explicitly analyze conditional specialization mechanisms (e.g., expert routing or format-/intent-dependent pathways) across heterogeneous datasets and question types.

To isolate the effect of \emph{routing} in a controlled comparison, we implement a \textbf{monolithic} baseline that mirrors the \textsc{CaReAQA}-style pipeline: (i) a \emph{single} audio encoder processes every spectrogram, (ii) a \emph{single} audio-to-LLM aligner produces aligned audio embeddings, and (iii) a \emph{single} LLM pathway (one adapter (one set of trainable parameters) on a frozen backbone, matching our training recipe) generates the answer from the prompt plus aligned audio prefix. This baseline keeps the multimodal interface (audio-prefix injection and autoregressive objective) comparable to RAMoEA-QA, while forcing all examples through one fixed acoustic representation and one fixed generation behavior.

%% file: figures/figures_tex/appendix_supplementary/pengi.tex
\begin{figure}[htbp]
\floatconts
  {fig:pengi}
  {\caption{Overview of the Pengi architecture used as a baseline in our experiments.}}
  {\includegraphics[width=0.8\linewidth]{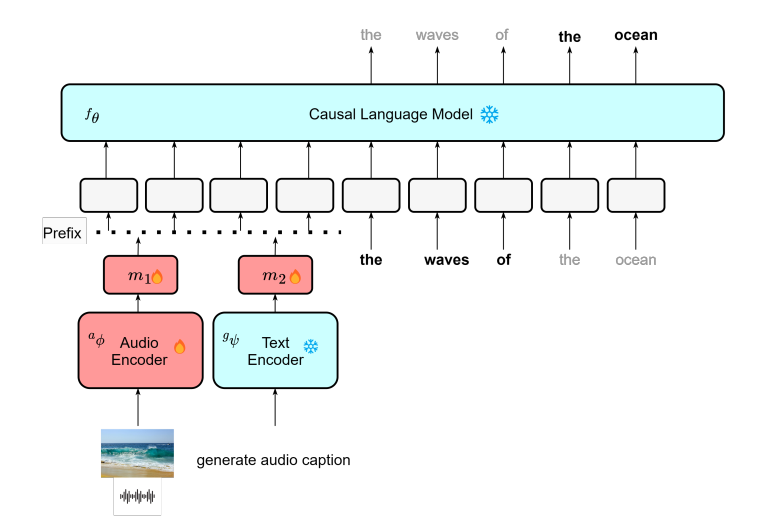}}
\end{figure}

%% file: appendix/example.tex
\section{Qualitative examples}
\label{apd:qualitative_examples}

To complement the aggregate quantitative results, we report two qualitative examples from the RA-QA benchmark: one \textbf{regression} example and one \textbf{discriminative} example. Together, they illustrate the different behavior of generic audio-language models, monolithic respiratory audio-QA baselines, and our routed model in clinically structured question answering.

\subsection{Regression example: respiratory-rate estimation}
\label{apd:qualitative_rr}

We first report a qualitative example from the \texttt{NoseMic} dataset, where the task is to estimate the patient's respiratory rate from audio. The input question is \emph{`What is the patient's respiratory rate?'} and the reference answer is \emph{`The patient's respiratory rate is 11.4 breaths/minute.'} Table~\ref{tab:qualitative_rr_example} compares the raw model outputs together with the parsed numeric prediction and the corresponding absolute error. For a single example, MAE and RMSE coincide for all parseable predictions, so we report only MAE.

\begin{table*}[t]
\centering
\scriptsize
\floatconts
  {tab:qualitative_rr_example}
  {\caption{\textbf{Qualitative regression example on respiratory-rate estimation.} \textbf{NA} indicates that no valid numeric value could be parsed from the generated answer.
  }}
{\begin{tabular}{lp{7.6cm}cc}
\toprule
\bfseries Model & \bfseries Prediction & \bfseries Parsed value & \bfseries MAE $\downarrow$ \\
\midrule
GAMA 
& The patient's respiratory rate is 24 breaths per minute. 
& 24.0 
& 12.60 \\ \midrule

LTU* 
& The patient's respiratory rate cannot be accurately determined based on the given audio clip alone. 
& NA 
& NA \\ \midrule

CareAQA-CT 
& The patient's respiratory rate is 12.22 breaths/minute. 
& 12.22 
& \underline{0.82} \\ \midrule

CareAQA-GT 
& The patient's respiratory rate is 12.22 breaths/minute. 
& 12.22 
& \underline{0.82} \\ \midrule

\textbf{RAMoEA-QA} 
& The patient's respiratory rate is 11.33 breaths/minute. 
& 11.33 
& \textbf{0.07} \\
\bottomrule
\end{tabular}}
\end{table*}

In this example, \textbf{RAMoEA-QA} produces the most accurate quantitative prediction, with a parsed value of \(11.33\) against the reference \(11.4\), corresponding to an absolute error of only \(0.07\). The two monolithic respiratory baselines, \textbf{CareAQA-CT} and \textbf{CareAQA-GT}, remain reasonably close but are less precise, both predicting \(12.22\) (MAE \(=0.82\)). By contrast, \textbf{GAMA} generates a fluent but substantially incorrect answer (\(24.0\), MAE \(=12.60\)), while \textbf{LTU} abstains from producing a numeric estimate altogether, stating that the respiratory rate cannot be determined from the audio alone.

This example reflects the broader pattern observed in the quantitative evaluation: RAMoEA-QA not only preserves the expected answer format, but also extracts more accurate clinically relevant numerical information than both generic audio-language models and comparable single-path respiratory baselines.

\subsection{Discriminative example: diagnosis prediction}
\label{apd:qualitative_dx}

We next report a qualitative example from the \texttt{KAUH} dataset, where the task is to identify the patient's diagnosis from respiratory audio. The input question is \emph{`What is the diagnosis for the patient?'} and the reference answer is \emph{`The diagnosis for the patient is a lower respiratory tract infection.'} Table~\ref{tab:qualitative_dx_example} compares the raw model outputs together with the corresponding task-level outcome after label parsing.

\begin{table*}[t]
\centering
\scriptsize
\floatconts
  {tab:qualitative_dx_example}
  {\caption{\textbf{Qualitative discriminative example on diagnosis prediction.} 
  }}
{\begin{tabular}{lp{7.6cm}cc}
\toprule
\bfseries Model & \bfseries Prediction & \bfseries Parsed label & \bfseries Correct \\
\midrule
GAMA 
& The patient has a severe case of pneumonia. 
& pneumonia 
& No \\ \midrule

LTU* 
& The diagnosis for the patient cannot be determined based on the given information. 
& NA 
& No \\ \midrule

CareAQA-CT 
& The diagnosis for the patient is chronic obstructive pulmonary disease. 
& COPD
& No \\ \midrule

CareAQA-GT 
& The diagnosis for the patient is chronic obstructive pulmonary disease. 
& COPD
& No \\ \midrule

\textbf{RAMoEA-QA} 
& The diagnosis for the patient is a lower respiratory tract infection. 
& LRTI
& \textbf{Yes} \\
\bottomrule
\end{tabular}}
\end{table*}

In this example, only \textbf{RAMoEA-QA} recovers the correct diagnosis. Both monolithic respiratory baselines predict \emph{chronic obstructive pulmonary disease}, which is clinically plausible as a respiratory condition but does not match the target label. \textbf{GAMA} instead generates \emph{pneumonia}, which is semantically related to the clinical scenario but still does not match the benchmark label after parsing. \textbf{LTU} abstains, stating that the diagnosis cannot be determined from the available information.

This example highlights an important aspect of structured clinical QA evaluation: responses may appear fluent or clinically plausible while still being task-incorrect after label normalization. In contrast, RAMoEA-QA not only generates a well-formed answer, but also aligns it with the expected diagnostic target. Together with the regression example above, this supports the broader conclusion that hierarchical routing improves both quantitative precision and label-level correctness in respiratory audio question answering.

%% file: appendix/results.tex
\section{Further results}
\label{apd:results}

This appendix provides a detailed breakdown of results for the RA-QA benchmark, covering both \textbf{discriminative} and \textbf{regression} tasks, as well as additional control and scaling experiments for the proposed framework. 

For discriminative tasks, we report per-dataset performance across question types (\emph{Global}, \emph{Single-verify}, \emph{Open-ended}, and \emph{Multiple-choice}) using standard classification and text-level metrics, including Accuracy, Macro F1, Token F1, Exact Match, BERTScore, and METEOR. For regression tasks, we report per-dataset performance for each target variable (e.g., MM Lung: FEV1/FVC; NoseMic: respiratory rate) using error-based metrics such as MAE and RMSE.

In addition to the full benchmark breakdown, this section includes supplementary analyses of the proposed framework, including single-stage and capacity-matched controls, a post-hoc ensemble comparison, and scaling experiments on the number of audio experts and language adapters. Together, these tables support comparison across datasets, question types, task families, and architectural variants, highlighting how performance varies both with the underlying data and with the design choices of the model.

\input{tables/appendix}
\input{tables/appendix_supplementary/operaCT_reg}
\input{tables/appendix_supplementary/operaGT_reg}
\input{tables/appendix_supplementary/operaCT_GT_reg}

\input{tables/appendix_supplementary/operaGT_ALL}

\input{tables/appendix_supplementary/operaCT_ALL}

\input{tables/appendix_supplementary/operaCT_GT_ALL}

%% file: tables/appendix.tex
\begin{table*}[t]
\centering
\scriptsize
\floatconts
  {tab:question_types}
  {\caption{\textbf{Per-question-type results for in-domain testing.} We report Macro F1 and BERTScore for each question type. \textbf{Bold} indicates the best result and \underline{underlining} indicates the second best.}
  }

{\begin{tabular}{lcc|cc|cc}
\toprule
\bfseries Model
& \multicolumn{2}{c}{\bfseries Single-Verify}
& \multicolumn{2}{c}{\bfseries Open-ended}
& \multicolumn{2}{c}{\bfseries Multiple-choice} \\
\midrule
& Macro F1 $\uparrow$ & BERTScore $\uparrow$
& Macro F1 $\uparrow$ & BERTScore $\uparrow$
& Macro F1 $\uparrow$ & BERTScore $\uparrow$ \\
\midrule
PENGI & 0.40 & 0.03 & 0.00 & -0.09 & 0.00 & -0.08\\
GAMA & 0.48 & 0.85 & 0.05 & 0.55 & 0.38 & 0.60 \\
\midrule
CareAQA-CT & 0.54 $\pm$ 0.10 & 0.87 $\pm$ 0.02 & \underline{0.49 $\pm$ 0.05} & \underline{0.86 $\pm$ 0.00} & \underline{0.59 $\pm$ 0.08} & \underline{0.85 $\pm$ 0.02} \\
CareAQA-GT & \underline{0.62 $\pm$ 0.28} & \underline{0.91 $\pm$ 0.01} & \textbf{0.50 $\pm$ 0.06} & \textbf{0.87 $\pm$ 0.00} & \underline{0.59 $\pm$ 0.08} & 0.84 $\pm$ 0.00 \\
\midrule
\textbf{RAMoEA-QA} & \textbf{0.82 $\pm$ 0.03} & \textbf{0.94 $\pm$ 0.00} & 0.42 $\pm$ 0.04 & 0.83 $\pm$ 0.02 & \textbf{0.61 $\pm$ 0.01} & \textbf{0.85 $\pm$ 0.03} \\
\bottomrule
\end{tabular}}
\end{table*}
\begin{table*}[t]
\centering
\scriptsize
\floatconts
  {tab:appendix_controls}
  {\caption{\textbf{Single-stage specialization controls and comparison against a larger single adapter.} We compare the full two-stage RAMoEA-QA model against single-stage variants and a capacity-increased single-adapter control. \textbf{Bold} indicates the best result and \underline{underlining} indicates the second best.}}
{\begin{tabular}{lccc}
\toprule
\bfseries Setting & Accuracy $\uparrow$ & MAE $\downarrow$ & BERTScore $\uparrow$ \\
\midrule
operaCT + 2 LoRA & 0.61 & 2.68 & 0.87 \\
operaGT + 2 LoRA & \underline{0.70} & 2.58 & \underline{0.89} \\
operaCT / operaGT + 1 LoRA & 0.68 & \underline{2.55} & 0.86 \\ \midrule
operaCT / operaGT + 1 LoRA (double rank) & 0.70 & \underline{2.55} & 0.88 \\ \midrule
\textbf{RAMoEA-QA} & \textbf{0.72} & \textbf{2.29} & \textbf{0.90} \\
\bottomrule
\end{tabular}}
\end{table*}

\begin{table*}[t]
\centering
\scriptsize
\floatconts
  {tab:appendix_scaling}
  {\caption{\textbf{Scaling the number of audio experts and language adapters.} We compare the default two-stage configuration against variants with a larger audio expert pool and increased numbers of LoRA adapters. \textbf{Bold} indicates the best result and \underline{underlining} indicates the second best.}}
{\begin{tabular}{lccc}
\toprule
\bfseries Setting & Accuracy $\uparrow$ & MAE $\downarrow$ & BERTScore $\uparrow$ \\
\midrule
operaCT / operaGT + 2 LoRA & 0.72 & \textbf{2.29} & 0.90 \\
operaCT / operaGT + 8 LoRA & \underline{0.75} & 2.79 & \underline{0.91} \\
operaCT / operaGT + 4 LoRA & 0.71 & 2.54 & \underline{0.91} \\
operaCT / operaGT / operaCE + 2 LoRA & \textbf{0.86} & \underline{2.55} & \textbf{0.94} \\
\bottomrule
\end{tabular}}
\end{table*}

\begin{table*}[t]
\centering
\scriptsize
\floatconts
  {tab:appendix_ensemble}
  {\caption{\textbf{Post-hoc ensemble control.} We compare a lightweight majority-vote baseline built from the strongest single-path models against the full RAMoEA-QA model. \textbf{Bold} indicates the best result.}}
{\begin{tabular}{lccc}
\toprule
\bfseries Model & Accuracy $\uparrow$ & Macro F1 $\uparrow$ & MAE $\downarrow$ \\
\midrule
Majority Vote & 0.61 & 0.54 & 2.62 \\ \midrule
\textbf{RAMoEA-QA} & \textbf{0.72} & \textbf{0.67} & \textbf{2.29} \\
\bottomrule
\end{tabular}}
\end{table*}

%% file: tables/appendix_supplementary/operaCT_reg.tex
\begin{table}[htbp]
\scriptsize
\centering
\caption{Per-dataset breakdown for in-domain testing (mean $\pm$ std over runs) for the baseline using operaCT on regressive tasks.}
\label{tab:reg_mmlung_nosemic_openended}
\begin{tabular}{ll|ccc}
\toprule
\bfseries Question-type & \bfseries Metric & \multicolumn{2}{c}{\bfseries MM Lung} & \bfseries NoseMic\\ \midrule
      && FEV1& FVC& Respiratory rate        \\ \midrule
\multirow{2}{*}{Open-ended} & MSE $\downarrow$     & $1.11 \pm 0.30$& $3.42 \pm 0.28$          & $27.79 \pm 2.34$        \\
      & RMSE $\downarrow$    & $1.05 \pm 0.14$& $1.85 \pm 0.08$          & $5.27 \pm 0.22$   \\
\bottomrule
\end{tabular}
\end{table}

%% file: tables/appendix_supplementary/operaGT_reg.tex
\begin{table}[htbp]
\scriptsize
\centering
\caption{Per-dataset breakdown for in-domain testing (mean $\pm$ std over runs) for the baseline using operaGT on regressive tasks.}
\label{tab:reg_mmlung_nosemic_openended}
\begin{tabular}{ll|ccc}
\toprule
\bfseries Question-type & \bfseries Metric & \multicolumn{2}{c}{\bfseries MM Lung} & \bfseries NoseMic\\ \midrule
      && FEV1& FVC& Respiratory rate        \\ \midrule
\multirow{2}{*}{Open-ended} & MSE $\downarrow$     & $1.13 \pm 0.28$& $2.73 \pm 0.69$          & $33.23 \pm 2.01$        \\
      & RMSE $\downarrow$    & $1.06 \pm 0.13$& $1.65 \pm 0.21$          & $5.76 \pm 0.17$   \\
\bottomrule
\end{tabular}
\end{table}

%% file: tables/appendix_supplementary/operaCT_GT_reg.tex
\begin{table}[htbp]
\scriptsize
\centering
\caption{Per-dataset breakdown for in-domain testing (mean $\pm$ std over runs) for RAMoEA-QA on regressive tasks.}
\label{tab:reg_mmlung_nosemic_openended}
\begin{tabular}{ll|ccc}
\toprule
\bfseries Question-type & \bfseries Metric & \multicolumn{2}{c}{\bfseries MM Lung} & \bfseries NoseMic\\ \midrule
      && FEV1& FVC& Respiratory rate        \\ \midrule
\multirow{2}{*}{Open-ended} & MSE $\downarrow$     & $0.96 \pm 0.13$& $2.42 \pm 1.00$          & $26.11 \pm 8.09$        \\
      & RMSE $\downarrow$    & $0.98 \pm 0.07$& $1.54 \pm 0.33$          & $5.08 \pm 0.80$   \\
\bottomrule
\end{tabular}
\end{table}

%% file: tables/appendix_supplementary/operaGT_ALL.tex
\begin{table*}[htbp]
\tiny
\floatconts
  {tab:operaGT_ALL}
  {\caption{Per-dataset breakdown for in-domain testing (mean $\pm$ std over runs) for the baseline configuration of operaGT on discriminative tasks.}}
{\begin{tabular}{ll|cccccc}
\toprule
\bfseries Metric & \bfseries Subset &
\multicolumn{1}{c}{\bfseries Global} &
\multicolumn{1}{c}{\bfseries Coswara} &
\multicolumn{1}{c}{\bfseries CoughVid} &
\multicolumn{1}{c}{\bfseries ICBHI} &
\multicolumn{1}{c}{\bfseries KAUH} &
\multicolumn{1}{c}{\bfseries Respiratory@TR}  \\
\midrule

\multirow{4}{*}{Accuracy $\uparrow$}
& Global          & $0.68 \pm 0.16$ & $0.75 \pm 0.35$ & $0.62 \pm 0.06$ & $0.50 \pm 0.00$ & $0.54 \pm 0.18$ & $0.88 \pm 0.04$ \\ \cmidrule(lr){2-8}
& Single-verify   & $0.70 \pm 0.21$ & $0.75 \pm 0.35$ & $0.67 \pm 0.00$ & $0.50 \pm 0.00$ & $0.67 \pm 0.24$ & $0.75 \pm 0.07$ \\
& Open-ended      & $0.60 \pm 0.06$ & -               & $0.33 \pm 0.00$ & $0.50 \pm 0.00$ & $0.50 \pm 0.24$ & $1.00 \pm 0.00$ \\
& Multiple-choice & $0.68 \pm 0.06$ & -               & $0.83 \pm 0.24$ & $0.50 \pm 0.00$ & $0.33 \pm 0.00$ & $1.00 \pm 0.00$ \\ \midrule

\multirow{4}{*}{Macro F1 $\uparrow$}
& Global          & $0.59 \pm 0.21$ & $0.67 \pm 0.47$ & $0.55 \pm 0.08$ & $0.33 \pm 0.00$ & $0.43 \pm 0.23$ & $0.87 \pm 0.05$ \\ \cmidrule(lr){2-8}
& Single-verify   & $0.62 \pm 0.28$ & $0.67 \pm 0.47$ & $0.62 \pm 0.00$ & $0.33 \pm 0.00$ & $0.58 \pm 0.35$ & $0.74 \pm 0.09$ \\
& Open-ended      & $0.50 \pm 0.06$ & -               & $0.18 \pm 0.02$ & $0.33 \pm 0.00$ & $0.40 \pm 0.22$ & $1.00 \pm 0.00$ \\
& Multiple-choice & $0.59 \pm 0.08$ & -               & $0.78 \pm 0.31$ & $0.33 \pm 0.00$ & $0.17 \pm 0.00$ & $1.00 \pm 0.00$ \\ \midrule

\multirow{4}{*}{Token F1 $\uparrow$}
& Global          & $0.86 \pm 0.06$ & $0.87 \pm 0.19$ & $0.90 \pm 0.02$ & $0.81 \pm 0.00$ & $0.85 \pm 0.05$ & $0.88 \pm 0.02$ \\ \cmidrule(lr){2-8}
& Single-verify   & $0.89 \pm 0.10$ & $0.87 \pm 0.19$ & $0.93 \pm 0.01$ & $0.82 \pm 0.00$ & $0.89 \pm 0.08$ & $0.97 \pm 0.01$ \\
& Open-ended      & $0.81 \pm 0.01$ & -               & $0.79 \pm 0.01$ & $0.79 \pm 0.00$ & $0.83 \pm 0.04$ & $0.83 \pm 0.00$ \\
& Multiple-choice & $0.82 \pm 0.01$ & -               & $0.97 \pm 0.05$ & $0.79 \pm 0.00$ & $0.79 \pm 0.00$ & $0.76 \pm 0.09$ \\ \midrule

\multirow{4}{*}{Exact Match $\uparrow$}
& Global          & $0.56 \pm 0.14$ & $0.75 \pm 0.35$ & $0.62 \pm 0.06$ & $0.50 \pm 0.00$ & $0.54 \pm 0.18$ & $0.47 \pm 0.04$ \\ \cmidrule(lr){2-8}
& Single-verify   & $0.70 \pm 0.21$ & $0.75 \pm 0.35$ & $0.67 \pm 0.00$ & $0.50 \pm 0.00$ & $0.67 \pm 0.24$ & $0.75 \pm 0.07$ \\
& Open-ended      & $0.38 \pm 0.06$ & -               & $0.33 \pm 0.00$ & $0.50 \pm 0.00$ & $0.50 \pm 0.24$ & $0.20 \pm 0.00$ \\
& Multiple-choice & $0.46 \pm 0.06$ & -               & $0.83 \pm 0.24$ & $0.50 \pm 0.00$ & $0.33 \pm 0.00$ & $0.20 \pm 0.00$ \\ \midrule

\multirow{4}{*}{BERTScore (F1) $\uparrow$}
& Global          & $0.89 \pm 0.00$ & $0.94 \pm 0.04$ & $0.86 \pm 0.03$ & $0.88 \pm 0.04$ & $0.90 \pm 0.01$ & $0.84 \pm 0.01$ \\ \cmidrule(lr){2-8}
& Single-verify   & $0.93 \pm 0.02$ & $0.94 \pm 0.04$ & $0.97 \pm 0.02$ & $0.93 \pm 0.07$ & $1.00 \pm 0.00$ & $0.82 \pm 0.01$ \\
& Open-ended      & $0.78 \pm 0.00$ & -               & $0.61 \pm 0.01$ & $0.83 \pm 0.00$ & $0.79 \pm 0.01$ & $0.87 \pm 0.01$ \\
& Multiple-choice & $0.85 \pm 0.02$ & -               & $0.89 \pm 0.08$ & $0.81 \pm 0.03$ & $0.81 \pm 0.03$ & $0.87 \pm 0.01$ \\ \midrule

\multirow{4}{*}{METEOR $\uparrow$}
& Global          & $0.88 \pm 0.01$ & $0.94 \pm 0.04$ & $0.86 \pm 0.05$ & $0.85 \pm 0.06$ & $0.89 \pm 0.01$ & $0.82 \pm 0.01$ \\ \cmidrule(lr){2-8}
& Single-verify   & $0.92 \pm 0.02$ & $0.94 \pm 0.04$ & $0.93 \pm 0.05$ & $0.93 \pm 0.08$ & $1.00 \pm 0.00$ & $0.81 \pm 0.01$ \\
& Open-ended      & $0.76 \pm 0.01$ & -               & $0.67 \pm 0.03$ & $0.77 \pm 0.03$ & $0.78 \pm 0.01$ & $0.83 \pm 0.01$ \\
& Multiple-choice & $0.82 \pm 0.02$ & -               & $0.91 \pm 0.08$ & $0.76 \pm 0.05$ & $0.79 \pm 0.03$ & $0.83 \pm 0.01$ \\

\bottomrule
\end{tabular}}
\end{table*}

%% file: tables/appendix_supplementary/operaCT_ALL.tex
\begin{table*}[htbp]
\tiny
\floatconts
  {tab:operaCT_ALL}
  {\caption{Per-dataset breakdown for in-domain testing (mean $\pm$ std over runs) for the baseline configuration of operaCT on discriminative tasks.}}
{\begin{tabular}{ll|cccccc}
\toprule
\bfseries Metric & \bfseries Subset &
\multicolumn{1}{c}{\bfseries Global} &
\multicolumn{1}{c}{\bfseries Coswara} &
\multicolumn{1}{c}{\bfseries CoughVid} &
\multicolumn{1}{c}{\bfseries ICBHI} &
\multicolumn{1}{c}{\bfseries KAUH} &
\multicolumn{1}{c}{\bfseries Respiratory@TR}  \\
\midrule

\multirow{4}{*}{Accuracy $\uparrow$}
& Global          & $0.61 \pm 0.02$ & $0.59 \pm 0.12$ & $0.65 \pm 0.14$ & $0.56 \pm 0.09$ & $0.47 \pm 0.04$ & $0.79 \pm 0.02$ \\ \cmidrule(lr){2-8}
& Single-verify  & $0.60 \pm 0.03$ & $0.59 \pm 0.12$ & $0.72 \pm 0.16$ & $0.62 \pm 0.18$ & $0.53 \pm 0.20$ & $0.58 \pm 0.04$ \\
& Open-ended     & $0.60 \pm 0.06$ &           -      & $0.33 \pm 0.00$ & $0.50 \pm 0.00$ & $0.50 \pm 0.24$ & $1.00 \pm 0.00$ \\
& Multiple-choice& $0.68 \pm 0.06$ &           -      & $0.83 \pm 0.24$ & $0.50 \pm 0.00$ & $0.33 \pm 0.00$ & $1.00 \pm 0.00$ \\ \midrule

\multirow{4}{*}{Macro F1 $\uparrow$}
& Global          & $0.54 \pm 0.06$ & $0.50 \pm 0.24$ & $0.58 \pm 0.18$ & $0.43 \pm 0.14$ & $0.40 \pm 0.03$ & $0.77 \pm 0.01$ \\ \cmidrule(lr){2-8}
& Single-verify  & $0.54 \pm 0.10$ & $0.50 \pm 0.24$ & $0.69 \pm 0.20$ & $0.53 \pm 0.28$ & $0.51 \pm 0.17$ & $0.54 \pm 0.02$ \\
& Open-ended     & $0.49 \pm 0.05$ &      -           & $0.17 \pm 0.00$ & $0.33 \pm 0.00$ & $0.40 \pm 0.22$ & $1.00 \pm 0.00$ \\
& Multiple-choice& $0.59 \pm 0.08$ &      -           & $0.78 \pm 0.31$ & $0.33 \pm 0.00$ & $0.17 \pm 0.00$ & $1.00 \pm 0.00$ \\ \midrule

\multirow{4}{*}{Token F1 $\uparrow$}
& Global          & $0.84 \pm 0.02$ & $0.78 \pm 0.07$ & $0.91 \pm 0.04$ & $0.83 \pm 0.03$ & $0.82 \pm 0.02$ & $0.89 \pm 0.00$ \\ \cmidrule(lr){2-8}
& Single-verify  & $0.84 \pm 0.03$ & $0.78 \pm 0.07$ & $0.93 \pm 0.05$ & $0.87 \pm 0.06$ & $0.83 \pm 0.06$ & $0.96 \pm 0.00$ \\
& Open-ended     & $0.81 \pm 0.01$ &       -          & $0.79 \pm 0.00$ & $0.79 \pm 0.00$ & $0.83 \pm 0.04$ & $0.83 \pm 0.00$ \\
& Multiple-choice& $0.84 \pm 0.01$ &         -        & $0.97 \pm 0.05$ & $0.79 \pm 0.00$ & $0.79 \pm 0.00$ & $0.83 \pm 0.00$ \\ \midrule

\multirow{4}{*}{Exact Match $\uparrow$}
& Global          & $0.50 \pm 0.02$ & $0.59 \pm 0.12$ & $0.65 \pm 0.14$ & $0.56 \pm 0.09$ & $0.47 \pm 0.04$ & $0.39 \pm 0.02$ \\ \cmidrule(lr){2-8}
& Single-verify  & $0.60 \pm 0.03$ & $0.59 \pm 0.12$ & $0.72 \pm 0.16$ & $0.62 \pm 0.18$ & $0.53 \pm 0.20$ & $0.58 \pm 0.04$ \\
& Open-ended     & $0.38 \pm 0.06$ &      -           & $0.33 \pm 0.00$ & $0.50 \pm 0.00$ & $0.50 \pm 0.24$ & $0.20 \pm 0.00$ \\
& Multiple-choice& $0.46 \pm 0.06$ &     -            & $0.83 \pm 0.24$ & $0.50 \pm 0.00$ & $0.33 \pm 0.00$ & $0.20 \pm 0.00$ \\ \midrule

\multirow{4}{*}{BERTScore (F1) $\uparrow$}
& Global          & $0.87 \pm 0.01$ & $0.82 \pm 0.07$ & $0.90 \pm 0.03$ & $0.87 \pm 0.03$ & $0.86 \pm 0.02$ & $0.89 \pm 0.02$ \\ \cmidrule(lr){2-8}
& Single-verify  & $0.90 \pm 0.02$ & $0.82 \pm 0.07$ & $0.96 \pm 0.03$ & $0.90 \pm 0.06$ & $0.89 \pm 0.04$ & $0.93 \pm 0.01$ \\
& Open-ended     & $0.82 \pm 0.01$ &      -           & $0.72 \pm 0.00$ & $0.83 \pm 0.00$ & $0.88 \pm 0.01$ & $0.85 \pm 0.03$ \\
& Multiple-choice& $0.86 \pm 0.02$ &      -           & $0.96 \pm 0.06$ & $0.83 \pm 0.00$ & $0.79 \pm 0.00$ & $0.85 \pm 0.03$ \\ \midrule

\multirow{4}{*}{METEOR $\uparrow$}
& Global          & $0.85 \pm 0.01$ & $0.81 \pm 0.06$ & $0.89 \pm 0.05$ & $0.84 \pm 0.03$ & $0.83 \pm 0.02$ & $0.88 \pm 0.01$ \\ \cmidrule(lr){2-8}
& Single-verify  & $0.88 \pm 0.02$ & $0.81 \pm 0.06$ & $0.91 \pm 0.07$ & $0.89 \pm 0.05$ & $0.86 \pm 0.06$ & $0.94 \pm 0.00$ \\
& Open-ended     & $0.80 \pm 0.01$ &      -           & $0.77 \pm 0.00$ & $0.79 \pm 0.02$ & $0.83 \pm 0.05$ & $0.81 \pm 0.01$ \\
& Multiple-choice& $0.83 \pm 0.02$ &      -           & $0.96 \pm 0.06$ & $0.77 \pm 0.00$ & $0.77 \pm 0.00$ & $0.81 \pm 0.01$ \\

\bottomrule
\end{tabular}}
\end{table*}

%% file: tables/appendix_supplementary/operaCT_GT_ALL.tex
\begin{table*}[htbp]
\tiny
\floatconts
  {tab:ramoeaqa_ALL}
  {\caption{Per-dataset breakdown for in-domain testing (mean $\pm$ std over runs) for RAMoEA-QA on discriminative tasks.}}
{\begin{tabular}{ll|cccccc}
\toprule
\bfseries Metric & \bfseries Subset &
\multicolumn{1}{c}{\bfseries Global} &
\multicolumn{1}{c}{\bfseries Coswara} &
\multicolumn{1}{c}{\bfseries CoughVid} &
\multicolumn{1}{c}{\bfseries ICBHI} &
\multicolumn{1}{c}{\bfseries KAUH} &
\multicolumn{1}{c}{\bfseries Respiratory@TR}  \\
\midrule

\multirow{4}{*}{Accuracy $\uparrow$}
& Global         & $0.72 \pm 0.02$ & $0.86 \pm 0.09$ & $0.54 \pm 0.14$ & $0.58 \pm 0.15$ & $0.68 \pm 0.02$ & $0.79 \pm 0.02$ \\ \cmidrule(lr){2-8}
& Single-verify  & $0.81 \pm 0.00$ & $0.86 \pm 0.09$ & $0.78 \pm 0.16$ & $0.72 \pm 0.22$ & $1.00 \pm 0.00$ & $0.57 \pm 0.04$ \\
& Open-ended     & $0.46 \pm 0.02$ & -               & $0.00 \pm 0.00$ & $0.44 \pm 0.09$ & $0.33 \pm 0.00$ & $1.00 \pm 0.00$ \\
& Multiple-choice& $0.62 \pm 0.10$ & -               & $0.61 \pm 0.24$ & $0.44 \pm 0.09$ & $0.39 \pm 0.08$ & $1.00 \pm 0.00$ \\ \midrule

\multirow{4}{*}{Macro F1 $\uparrow$}
& Global         & $0.67 \pm 0.03$ & $0.86 \pm 0.09$ & $0.51 \pm 0.18$ & $0.47 \pm 0.18$ & $0.60 \pm 0.02$ & $0.75 \pm 0.01$ \\ \cmidrule(lr){2-8}
& Single-verify  & $0.79 \pm 0.01$ & $0.86 \pm 0.09$ & $0.76 \pm 0.19$ & $0.67 \pm 0.29$ & $1.00 \pm 0.00$ & $0.51 \pm 0.03$ \\
& Open-ended     & $0.38 \pm 0.02$ & -               & $0.00 \pm 0.00$ & $0.26 \pm 0.11$ & $0.17 \pm 0.00$ & $1.00 \pm 0.00$ \\
& Multiple-choice& $0.53 \pm 0.12$ & -               & $0.51 \pm 0.34$ & $0.30 \pm 0.04$ & $0.23 \pm 0.09$ & $1.00 \pm 0.00$ \\ \midrule

\multirow{4}{*}{Token F1 $\uparrow$}
& Global         & $0.88 \pm 0.00$ & $0.93 \pm 0.05$ & $0.88 \pm 0.04$ & $0.84 \pm 0.06$ & $0.90 \pm 0.00$ & $0.83 \pm 0.01$ \\ \cmidrule(lr){2-8}
& Single-verify  & $0.92 \pm 0.01$ & $0.93 \pm 0.05$ & $0.95 \pm 0.04$ & $0.90 \pm 0.08$ & $1.00 \pm 0.00$ & $0.83 \pm 0.01$ \\
& Open-ended     & $0.78 \pm 0.00$ & -               & $0.71 \pm 0.02$ & $0.77 \pm 0.02$ & $0.79 \pm 0.00$ & $0.84 \pm 0.01$ \\
& Multiple-choice& $0.83 \pm 0.02$ & -               & $0.93 \pm 0.06$ & $0.76 \pm 0.04$ & $0.80 \pm 0.02$ & $0.84 \pm 0.01$ \\ \midrule

\multirow{4}{*}{Exact Match $\uparrow$}
& Global         & $0.58 \pm 0.02$ & $0.86 \pm 0.09$ & $0.54 \pm 0.14$ & $0.58 \pm 0.15$ & $0.68 \pm 0.02$ & $0.34 \pm 0.02$ \\ \cmidrule(lr){2-8}
& Single-verify  & $0.79 \pm 0.00$ & $0.86 \pm 0.09$ & $0.78 \pm 0.16$ & $0.72 \pm 0.22$ & $1.00 \pm 0.00$ & $0.47 \pm 0.04$ \\
& Open-ended     & $0.24 \pm 0.02$ & -               & $0.00 \pm 0.00$ & $0.44 \pm 0.09$ & $0.33 \pm 0.00$ & $0.20 \pm 0.00$ \\
& Multiple-choice& $0.40 \pm 0.10$ & -               & $0.61 \pm 0.24$ & $0.44 \pm 0.09$ & $0.39 \pm 0.08$ & $0.20 \pm 0.00$ \\ \midrule

\multirow{4}{*}{BERTScore (F1) $\uparrow$}
& Global         & $0.89 \pm 0.00$ & $0.94 \pm 0.04$ & $0.86 \pm 0.03$ & $0.88 \pm 0.04$ & $0.90 \pm 0.01$ & $0.84 \pm 0.01$ \\ \cmidrule(lr){2-8}
& Single-verify  & $0.93 \pm 0.02$ & $0.94 \pm 0.04$ & $0.97 \pm 0.02$ & $0.93 \pm 0.07$ & $1.00 \pm 0.00$ & $0.82 \pm 0.01$ \\
& Open-ended     & $0.78 \pm 0.00$ & -               & $0.61 \pm 0.01$ & $0.83 \pm 0.00$ & $0.79 \pm 0.01$ & $0.87 \pm 0.01$ \\
& Multiple-choice& $0.85 \pm 0.02$ & -               & $0.89 \pm 0.08$ & $0.81 \pm 0.03$ & $0.81 \pm 0.03$ & $0.87 \pm 0.01$ \\ \midrule

\multirow{4}{*}{METEOR $\uparrow$}
& Global         & $0.88 \pm 0.01$ & $0.94 \pm 0.04$ & $0.86 \pm 0.05$ & $0.85 \pm 0.06$ & $0.89 \pm 0.01$ & $0.82 \pm 0.01$ \\ \cmidrule(lr){2-8}
& Single-verify  & $0.92 \pm 0.02$ & $0.94 \pm 0.04$ & $0.93 \pm 0.05$ & $0.93 \pm 0.08$ & $1.00 \pm 0.00$ & $0.81 \pm 0.01$ \\
& Open-ended     & $0.76 \pm 0.01$ & -               & $0.67 \pm 0.03$ & $0.77 \pm 0.03$ & $0.78 \pm 0.01$ & $0.83 \pm 0.01$ \\
& Multiple-choice& $0.82 \pm 0.02$ & -               & $0.91 \pm 0.08$ & $0.76 \pm 0.05$ & $0.79 \pm 0.03$ & $0.83 \pm 0.01$ \\

\bottomrule
\end{tabular}}
\end{table*}